\documentclass[10pt,smallextended]{svjour3}       

\usepackage{mathptmx,natbib,amsmath,amsbsy,amssymb,amsfonts}      
\usepackage{figpaths,epsfig,epic,eepic,color,graphicx,wrapfig} 
\usepackage{enumitem,appendix,wrapfig}
\usepackage{geometry} 
\usepackage{setspace,subfigure,bm,float,lscape}
 \usepackage{esint}
\RequirePackage{fix-cm}
\smartqed  

\bibpunct{(}{)}{,}{a}{}{;} 

 \journalname{J Math Biol}

\numberwithin{equation}{section}
\numberwithin{figure}{section}

\newcommand{\bphi}{{\mbox{\bm{$\phi$}}}}
\newcommand{\bfe}{{\bm{e}}}
\newcommand{\bfu}{{\bm{u}}}
\newcommand{\bff}{{\mbox{\bm{$f$}}}}

\newcommand{\stress}{\mbox{$\bm{T}$}}  
\newcommand{\sig}{\mbox{$\bm{\sigma}$}}  

\newcommand{\bfeps}{{\bm \epsilon}}

\def\cO{{\mathcal{O}}}
\def\cF{{\mathcal F}}

\def\bfn{{\mathbf{n}}}
\def\cf{{\em cf.\,}}

\def\cS{{\mathcal S}}
\def\bfx{{\mathbf{x}}}

\def\bfu{{\mathbf{u}}}

\def\bF{{\mathbf{F}}}

\def\da{{{\,\rm d} A }}
\def\d2u{{{\rm d}^2u}}

\begin{document}

\title{Getting in shape and  swimming: the role of  cortical forces  and
  membrane heterogeneity in eukaryotic cells
\thanks{Supported in part by NSF Grants DMS 0817529 and 1311974, NIH Grant 54-CA-210190, NIH Grant GM29123, the Newton  Institute  and  the Simons Foundation.}}

\titlerunning{Eukaryotic cell swimming}        

\author{Hao Wu, Marco Avila Ponce de Le\'on and Hans G. Othmer  
}
\authorrunning{Wu, Ponce de Leon, Othmer} 

\institute{School of Mathematics, 270A Vincent Hall \\
              University of Minnesota\\
              Tel.: (612) 624-8325\\
              Fax: (612) 626-2017\\
              \email{wuxx1798@umn.edu,~~ponce018@umn.edu,~~othmer@math.umn.edu}      
}

\date{\today}

\bigskip

\maketitle


\bigskip
\begin{abstract}
Recent research has shown that motile cells can adapt their mode of propulsion to the mechanical properties of the environment in which they find themselves--crawling in some environments while swimming in others. The latter can involve movement by blebbing or other cyclic shape changes, and both highlysimplified and more realistic models of these modes have been studied previously. Herein we study swimming that is driven by membrane tension gradients that arise from flows in the actin cortex underlying the membrane, and does not involve imposed cyclic shape changes. Such gradients can lead to a number of different characteristic cell shapes, and our first objective is to understand how different distributions of membrane tension influence the shape of cells in an inviscid quiescent fluid. We then analyze the effects of spatial variation in other membrane properties, and how they interact with tension gradients to determine the shape. We also study the effect of fluid--cell interactions and show how tension leads to cell movement, how the balance between tension gradients and a variable bending modulus determine the shape and direction of movement, and how the efficiency of movement depends on the properties of the fluid and the distribution of tension and bending modulus in the membrane.

\keywords{Low Reynolds number swimming \and Self-propulsion
\and Membrane tension gradients \and Heterogeneous membrane \and Boundary
  integral method }
\subclass{49Q10 \and  49S05 \and 70G45 \and 92C10}
\bigskip
{\em Dedicated to the memory of Karl P. Hadeler, a pioneer in the field of
  Mathematical Biology and a friend and mentor to many.}

\end{abstract}

\section{Introduction}

Movement of cells, either individually or  collectively, plays an important role
in numerous biological processes, including development, the immune response,
wound healing, and cancer metastasis \citep{Nurnberg:2011:NAI}.  Single-cell
organisms exhibit a variety of modes for translocation, including crawling,
swimming, or drifting with a fluid flow.  Some prokaryotes such as bacteria use
flagella, while eukaryotes such as paramecia use cilia to swim, but both types
can only use one mode. However other eukaryotes, such as tumor cells, are more
flexible and can adopt the mode used to the environment in which they find
themselves.  For instance, whether a single-cell or collective mode of movement
is used in tissues can depend on the density of the 3D extracellular matrix (ECM)
in which cells find themselves \citep{Haeger:2015:CCM}.  This adaptability has
significant implications for developing new treatment protocols for cancer and
other diseases, for it implies that it is essential to understand the processes
by which cells detect extracellular chemical and mechanical signals and
transduce them into intracellular signals that lead to force generation,
morphological changes and directed movement. 

\begin{wrapfigure}[]{r}{2.75in}
\vspace*{-10pt}
\centering
\includegraphics[width=2.75in,height=2.in]{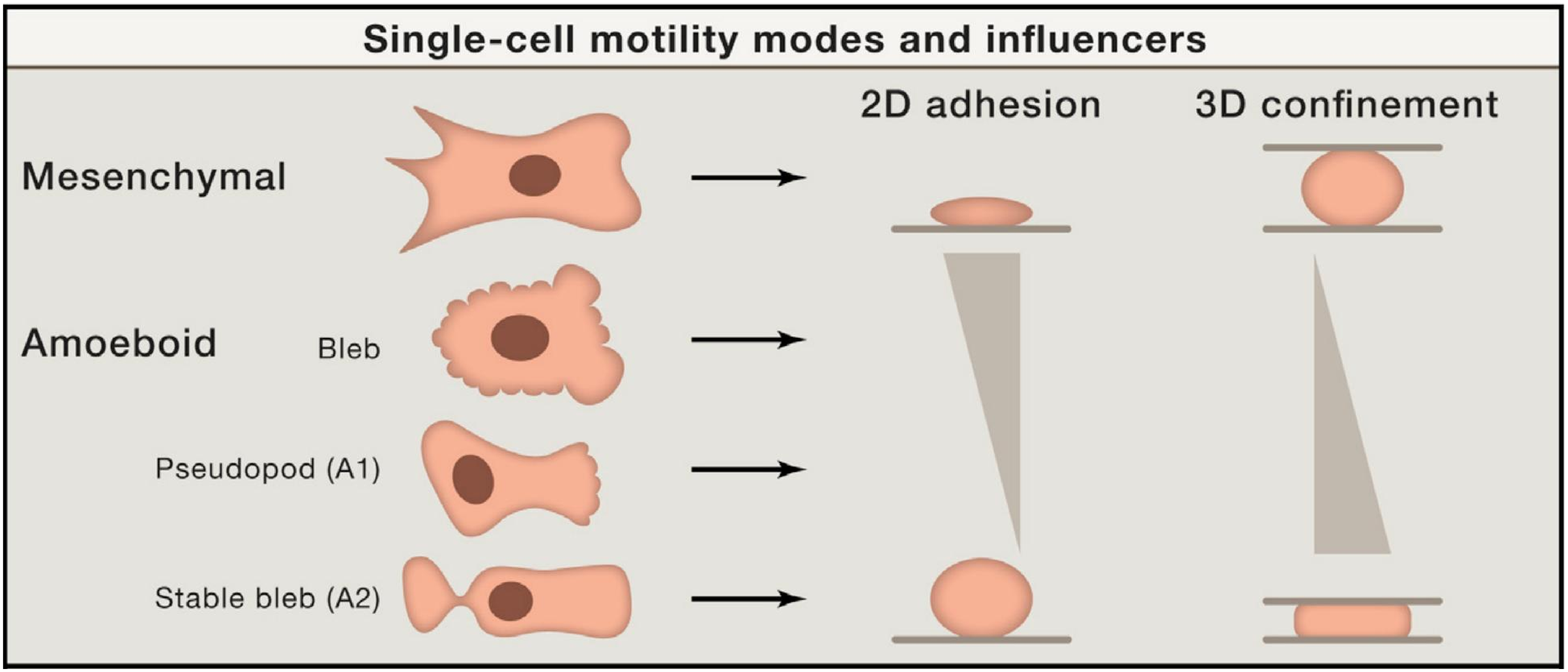}
\vspace*{-10pt}
\caption{  A summary of the
 different modes of movement in different environments and under different
 substrate properties. Modified from \protect \citet{Welch:2015:CMF}. }
\vspace*{-15pt}
\label{modes}
\end{wrapfigure}

The two primary modes of eukaryotic cell movement on surfaces or in the ECM are
called mesenchymal and amoeboid ({\it cf.}  Figure \ref{modes})
\citep{Friedl:2010:PCM,Biname:2010:WMC}.  The former is used by fibroblasts and
tumor cells of epithelial origin, and typically involves strong adhesion to the
substrate.  In 2D movement is by extension of relatively-flat lamellipodia at
the leading edge, whose protrusion is driven by actin polymerization. 
 Growth of this structure is understood in terms of the dendritic network hypothesis, which
posits that filaments are nucleated at the membrane and treadmill as in
solution, and that the densely-branched structure of the network arises via
nucleation of branches on existing filaments mediated by a protein called Arp2/3
\citep{Pollard:2000:MMC}.  Force is transmitted to the environment via
integrin-mediated focal adhesions that are connected to the internal network
(the cytoskeleton or CSK) via stress fibers. Movement frequently involves
proteolysis of the ECM to create a pathway \citep{Sanz-Moreno:2008:RAI}.
\begin{wrapfigure}[]{r}{2.5in}
\centering
\psfig{figure=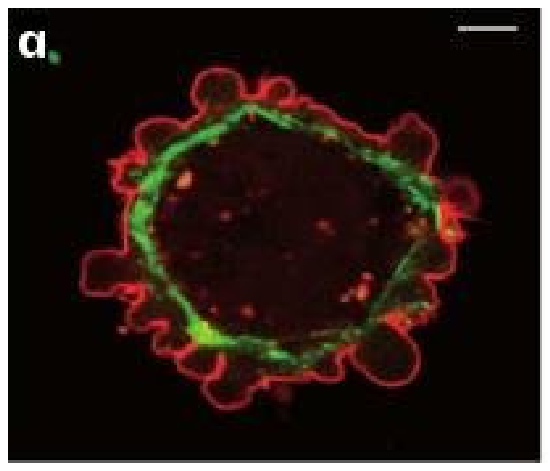,width=1.3in}\hspace*{3pt}\psfig{figure=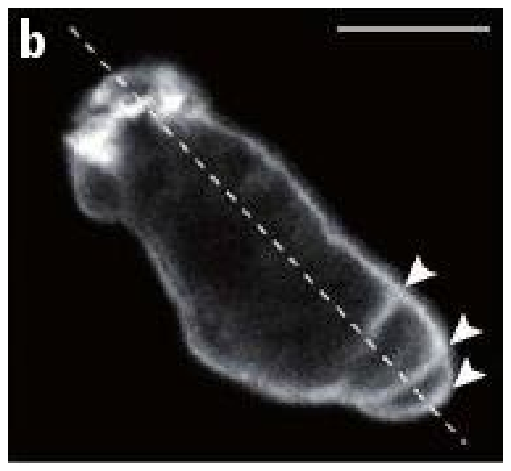,width=1.2in}
\vspace*{-10pt}
\caption{ (a) Blebbing on a melanoma cell: myosin (green) localizes under the
  blebbing membrane (red) (b) The actin cortex of a {\em Dictyostelium
    discoideum} (Dd) cell migrating to the lower right. Arrowheads indicate the
  successive blebs and arcs of the actin cortex \protect\citep{Charras:2008:BLW}.
}
\label{blebbing}
\end{wrapfigure}

In contrast, the amoeboid mode utilizes a less-structured CSK that lacks stress
fibers, and involves lower adhesion to the substrate. Proteolytic degradation of
the ECM is generally not used, and cells adopt a more rounded cell shape, often
with a highly contractile 'tail' called the uropod
\citep{Lammermann:2008:RLM}. Two sub-types of amoeboid motion are known.  In
one, cells move by generating a rearward flow in the cortex -- the cross-linked
filamentous actin (F-actin) network that is linked to the membrane. As described
later, the drag force created by the rearward flow leads to a reactive tension
gradient in the membrane that propels the cell forward, and this is called the
'tension-' or 'friction-driven' mode.  In another type, cells move by blebbing,
in which cycles of extension of the front and retraction of the rear as shown in
Figure \ref{blebbing}(b) are used. How the spatial localization of blebs shown
there is controlled is not understood, and some cells exhibit random blebs over
their surface ({\it cf.} Figure \ref{blebbing}(a)), which leads to no net
translocation. 
 Another variation of 
blebbing called 'stable-bleb' or 'leader-bleb' migration is used by certain
embryonic cells that form a balloon-like protrusion at their leading edge
(cf.~Figure \ref{modes}) and can move rapidly
\citep{Maiuri:2015:AFM,Ruprecht:2015:CCT,Logue:2015:ERA}.

The amoeboid mode is widely used, and when the environment is less favorable to
mesenchymal movement, due {\em e.g.}, to changes in the adhesiveness of the
substrate, cells compensate by undergoing a `mesenchymal-to-amoeboid' transition
(MAT) \citep{Friedl:2011:CIM,vanZijl:2011:ISM}.  Leukocytes can use the
mesenchymal mode in the ECM, but can also migrate {\em in vivo} in the absence
of integrins, using a 'flowing and squeezing' mechanism
\citep{Lammermann:2008:RLM}.  Dd moves in a cyclic AMP (cAMP) gradient either by
extending pseudopodia or by blebbing, and cells monitor the stiffness of the
surroundings to determine the mode: pseudopodia in a compliant medium and
blebbing in stiffer media \citep{Zatulovskiy:2014:BDC}.  Finally, some cells
move only by blebbing. Certain lines of carcinoma cells don't move on 2D
substrates, but in a confined environment they polarize spontaneously, form
blebs, and move efficiently \citep{Bergert:2012:CMC}.

A third, less-studied mode of movement is by unconfined swimming in a fluid. Dd
cells and neutrophils both do this \citep{Barry:2010:DAN}, presumably to move
through fluid-filled voids in their environment.  A model of swimming by shape
changes has been analyzed in \citep{Wang:2015:CAA}, where it is shown that Dd
can swim by propagating protrusions axially. The model gives insights into how
characteristics of the protrusions such as their height affect the swimmer's
speed and efficiency.  Simplified models for movement by repetitive blebbing,
using what can be described as a 'push-pull' mechanism, have been analyzed and the
efficiency of movement determined \citep{Wang:2015:PDM,Wang:2016:AMM}. However,
it is also observed that Dd cells can swim without shape changes or blebbing for several
body lengths \citep{Howe:2013:HDA}, and the authors state that 'Simply put, we
do not understand how these cells swim, and therefore how they move.' Herein we
show how swimming can arise by maintaining an axial tension gradient in the
membrane, and this would be difficult to detect at the macroscopic level.

Crawling and swimming are the extremes on a continuum of strategies, and the
variety of modes used in different environments raises questions about how
mechano-chemical sensing of the environment is used to control the evolution of
the CSK \citep{Renkawitz:2010:MFG}. Protrusions and other shape changes require
forces that must be correctly orchestrated in space and time to produce net
motion -- those on cells in Figure\,\ref{blebbing} (a) are not, while those in (b)
are -- and to understand this orchestration one must couple the intracellular
dynamics with the state of the surrounding fluid or ECM. Tension in the membrane
and cortex has emerged as an important determinant in the orchestration, whether
in the context of undirected cell movement, or in movement in response to
environmental cues. Experimental observations on several modes of 
 tension-driven movement are described in the next section.

\section{The basis of tension-driven movement} 

To describe recent experimental results we introduce some terminology. We
consider a cell as a three-layered structure comprised of the plasma membrane,
the cortex, and the remainder (cytosol, nucleus and other components). The
membrane is a lipid bilayer $\sim\negmedspace10$ nm thick, and the cortex, which
is 200-300 nm thick, is composed of an F-actin network cross-linked by filamin
and bound to the motor protein myosin-II (myo-II), which can confer rigidity to
the cortex, but can also contract and exert tension in the cortex.
Membrane-bound proteins such as myo-I -- a small motor protein that binds to
both actin and the membrane \citep{Dai:1999:MCG} -- or linker proteins such as
ERMs (ezrin, radixin, and moesin), tether the cortex to the membrane and exert a
normal force on the membrane.  However the connection is dynamic and when the
cortex flows it can slide tangentially under the membrane
\citep{Hochmuth:1996:DFM,Dai:1999:MTF} and exert a tangential stress on the
membrane. A detailed force balance done later shows how this can lead to
swimming without either blebbing or shape changes.

Two recent papers describe a similar phenomenon -- tension-driven motion of
 cells in confined environments -- using human dermal fibroblast cells
 \citep{Liu:2015:CLA}, or zebrafish germ-layer cells \citep{Ruprecht:2015:CCT}.
 \citet{Liu:2015:CLA} identify two morphologies, one type -- A1 -- has a rounded
 body and a small leading edge, and the other -- A2 -- has a more ellipsoidal
 body with a large uropod (Figure\,\ref{modes}). They showed that slow
 mesenchymal cells undergo the MAT when the adhesion is low and the cells are
 confined between plates, and this leads to two distinct shapes and two types of
 fast migration.
\begin{wrapfigure}[]{r}{2.25in}
\vspace*{-20pt}
\begin{center}
            \includegraphics[height=3.25cm]{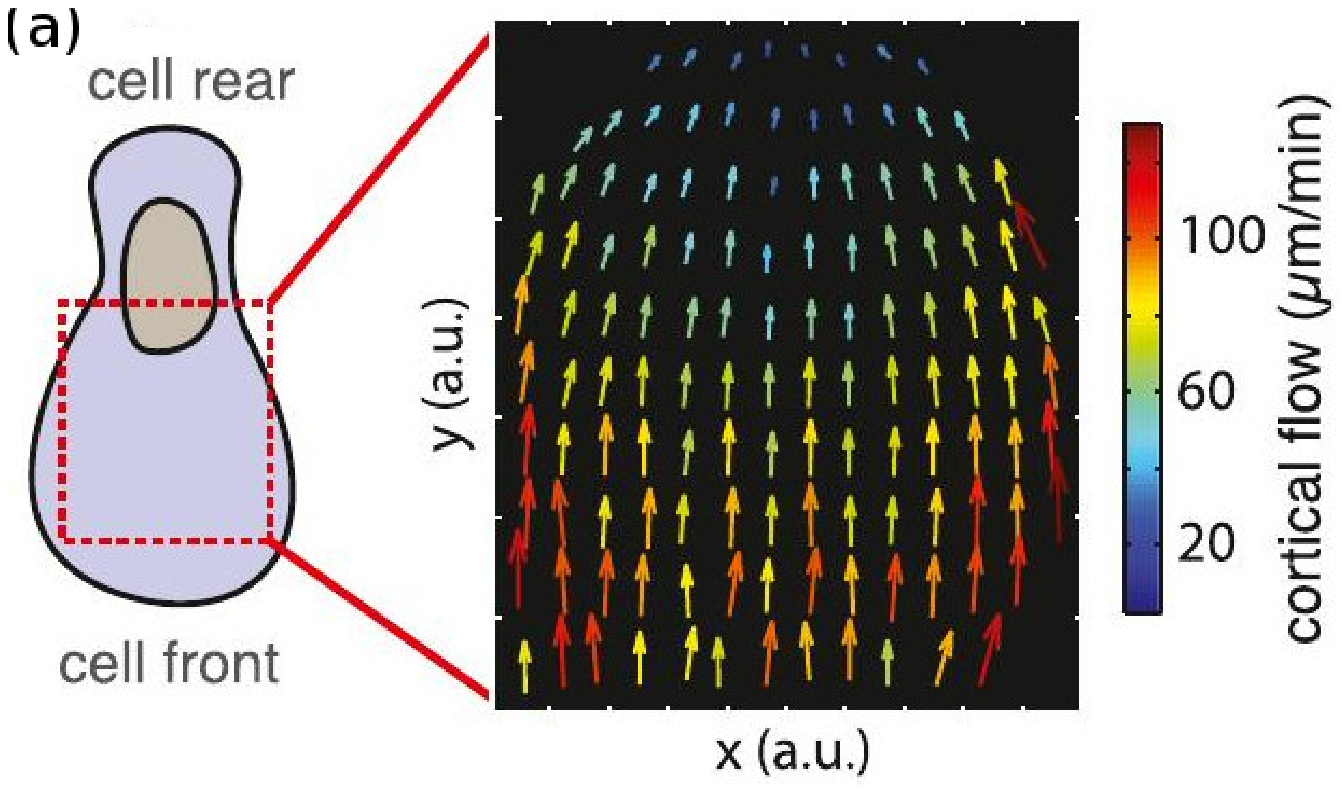}\\
\vspace*{10pt}
\includegraphics[width=5cm,height=5cm]{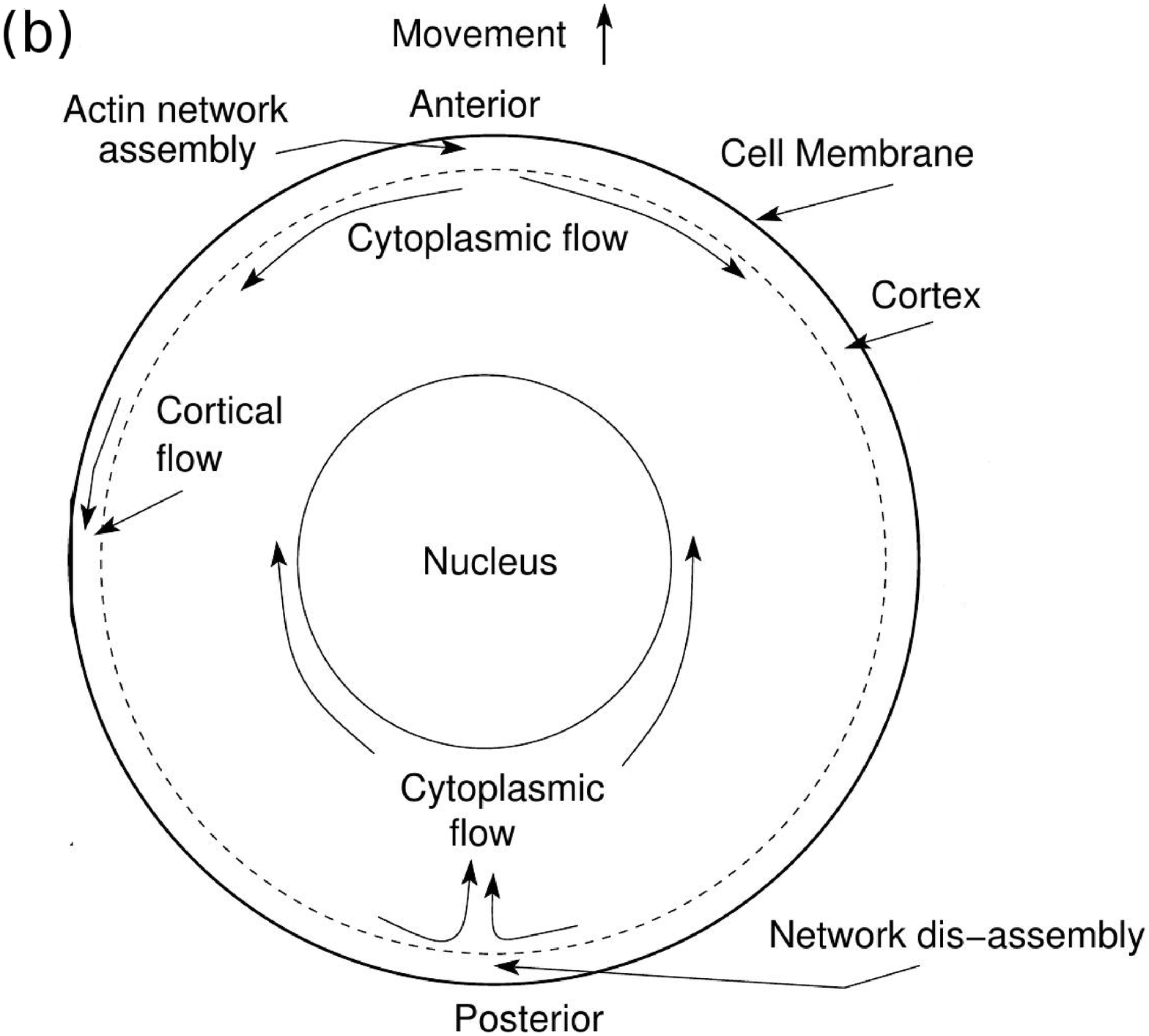} 
\end{center}  
\vspace*{-5pt}       
\caption{The measured cortical flow (top) (From \protect Ruprecht, et al. \citet{Ruprecht:2015:CCT}), and the postulated
  intracellular flows (bottom). }
\label{cortflow}
\vspace*{-18pt}
\end{wrapfigure}
 The first (A1) involves low contractility of the cortex and a
 local protrusion, and the second (A2), is a 'stable-bleb' type that involves high
 myo-II activity and involves a strong retrograde actin flow.  Type A1 appears
 to require an external signal to polarize, whereas type A2 can appear
 spontaneously, as has been shown for other cell types as well 
 \citep{Lorentzen:2011:ERR,Bergert:2012:CMC}.  The authors suggest that the
 type A2 system may be bistable.

 A different, permanently-polarized, stable-bleb shape can be obtained from a
 stable non-polarized blebbing cell by increasing the contractility in zebrafish
 progenitor cells \citep{Ruprecht:2015:CCT}.  The stable-bleb form involves
 cortical flow rates of 10's of $\mu$ms/min (Figure\,\ref{cortflow}(a)), which
 would certainly induce an anterior-to-posterior cytoplasmic flow near the
 cortex, and thus a posterior-to-anterior flow in the center.  The authors also
 postulate a high cortical growth rate at the front of a cell and a high
 disassembly rate at the rear (\cf Figure \ref{cortflow}(b)). This hypothesis is
 supported by the fact that blebbistan (which inhibits myo-II contractility),
 latrunculin A (which leads to actin depolymerization), and jasplakinolide
 (which stabilizes actin networks) all inhibit polarization and cortical
 flow. The authors explain the transition from random blebbing to a stable-bleb
 shape as an instability of a spherical shape, which is adopted in the absence
 of surface contact, to fluctuations in the membrane. To date only a linear
 stability analysis of the problem has been done \citep{Callan-Jones:2016:CFS},
 and the results of that analysis are contrasted with the results herein in the
 Discussion section.

\begin{wrapfigure}[]{r}{1.5in}
\vspace*{-10pt}
            \includegraphics[width=1.5in]{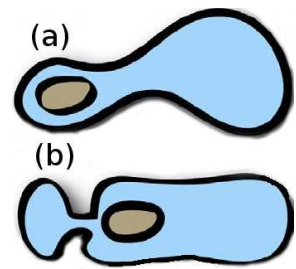}                                
\caption{The two types of  stable-bleb morphology in which movement in
  confinement  is driven by tension gradients. (a) The Ruprecht-type, and (b)
  the Liu-type A2. Reproduced with permission from \citep{Petrie2016multiple}.}
\label{2morph}
\vspace*{-20pt}
\end{wrapfigure}

Thus there are two 'stable-bleb' cell morphologies in which a gradient of
cortical density and myo-II contraction is used to generate a cortical flow and
an axial pressure gradient (\cf Figure \ref{2morph}).  How the flow is
initiated, and in particular, whether it arises as an instability or requires
contact with a substrate, is unknown. This mode of movement has only been
demonstrated when cells are constrained to move between two horizontal barriers,
and it has been suggested that substrate contact may open stretch-activated
calcium channels and initiate modification of the
cortex \citep{Hung:2016:CSS}. Interestingly, some cells cannot move if they are
only in contact with the substrate on the ventral (bottom) side, but will move
when confined in a micro-channel \citep{Bergert:2012:CMC}. This suggests that
cortical flow may not arise when the cell is in contact with a surface on only
one side, and this has been shown experimentally in Dd \citep{Yumura:2012:CDR}.

To understand the origin of movement in these cells, recall that the cortex
slips past the membrane, and in a numerous cell types, including Dd
\citep{Traynor:2007:PRE}, leukocytes \citep{Lee:1990:DML} and dendritic cells
\citep{Renkawitz:2009:AFT}, the membrane does not flow in a cell-fixed
coordinate frame -- it merely translocates with the cell body. Thus the membrane
functions more like an elastic than a viscous material, and the drag force due
to cortical flow creates an opposed tension gradient in the membrane.
Back-to-front tension gradients of the order of 5$pN/\mu m^2$ have been measured
in axons and keratocytes \citep{Lieber:2015:FRM}, but only on the dorsal
membrane of cells in surface contact on the ventral membrane.  When the internal
and external fluids are inviscid and quiescent, as in sections 3 and 4, these
tensions must be equal and opposite, and the cell does not translocate. However,
when the interior and exterior fluids are viscous, as in Section 5, there can be
a non-zero tangential stress on these fluids, and this may produce motion. We will
show that such tension gradients in the membrane can generate movement without
shape changes, and we call this 'swimming'.  The plausibility of such motion is
supported by the observed surface-tension-driven Marangoni propulsion of a fluid
droplet on a viscous fluid \citep{Lauga:2012:VMP}. In fact, by cutting the cell
in Figure \ref{cortflow} along the long axis, and opening it up and stretching
the free ends to infinity, the membrane becomes an interface between the
interior and exterior fluids, and one has the configuration of the classical
Marangoni problem. Of course reality is more complex here, since the cortex as
described above is a dynamic structure, thin at the front and thick at the rear,

While the primary experimental evidence of tension-driven movement involves
cortical flows, we show that it is the existence of a tension gradient, whether
or not it involves a cortical flow, that drives movement. Thus the results
herein will be applicable to a larger class of cells than those used in the
experiments described above.

\section{The shape problem for cells}

\subsection{The free energy functional and the shape equations} 

The observed shapes described above, in particular the stable-bleb types shown
in Figure \ref{2morph}, raise a number of questions.  Since amoeboid cells have
a less-structured CSK, the cell shape is  primarily determined  by the 
distribution of internal forces in the membrane and the forces in the
cortex. Thus the first question is what distribution of forces in the membrane
and the cortex is needed to produce the observed shapes? Secondly, since
directed cell movement requires cell polarization and cortical flow, what
balances in the cortex amongst cortical thickness, the level of myo-II for
contraction, myo-I for attachment to the membrane, and other factors, are needed
to produce the cortical flow?  Finally, how do the properties of the
micro-environment affect the speed of movement, and whether a cell switches from 
tension-driven movement  to blebbing-driven  movement? 

A model that incorporates a detailed description of the cortical network growth,
myo-II, and cortical-network interactions, combined with the transport of actin
monomers and other components in the CSK, will be very complex, and to date the
cortex has been described as an active gel in simplified treatments of cortical
flow \citep{Joanny:2009:AGD,Prost:2015:AGP,Bergert:2015:FTD}. While this produces
some insights, the biochemical details are embedded in an active component of
the stress tensor for the gel, and thus the relative importance of the
individual processes mentioned above cannot be investigated. We also do not
attempt to develop a detailed model here, but rather, we use an alternate
high-level description of the cortex to investigate how cortical forces and
heterogeneity of membrane properties determine the shape of cells in quiescent
fluids, and how these factors determine the shape and speed of swimming
cells. Since the forces are simply specified, we can consider both the case in
which there is a cortical flow that generates stresses, and the case in which
the cortex is under stress but not flowing with this approach.

For this purpose we separate mechanics from the details of the cortical
structure by prescribing a force distribution on the membrane that reflects what
is believed to occur in the cortex {\em in vivo}. This allows us to vary the
cortical forces directly, whether they arise from cortical flows or simply from
static tension gradients in the cortex. We do this first for cells in a
quiescent fluid, where there are no shear stresses due to the fluids, and
secondly for swimming cells, in which the interior and exterior fluids flow
freely. The three-fold aim of the latter step is to show that cells can swim
when subject to cortical forces, to determine how much the shape of a moving
cell differs from that of a stationary cell, and to show how the shape
differences depend on cell and fluid properties.

The determination of the steady-state shapes of vesicles and red blood cells has
been thoroughly studied, both in the absence of fluid motion
\citep{Seifert:1991:STV,Seifert:1997:CFM}, and in the presence of fluid motion
\citep{Veerapaneni:2009:NMF,Zhao:2010:SBI,Li:2014:PRB,Kaoui:2009:WRB,Ghigliotti:2010:RDT,Thiebaud:2013:RVS}. In the former the shapes
are computed as minimizers of the free energy of the membrane, typically given
by a Canham-Helfrich functional described below, and in the latter they
represent shapes that lead to minimum dissipation in the flow. However, vesicles
have no cortical layer and red blood cells have a very thin layer of spectrin --
which contains no molecular motors -- attached to the membrane.  When there is a
cortical flow and membrane-cortex tethers are actively formed and broken, there
are dissipative processes involved and the membrane-cortex forces are not
conservative.  While one can use a virtual work argument to determine stationary
shapes when there are non-conservative forces, we avoid this as follows. To find
the stationary shapes under prescribed forces, we define a free-energy
functional for the membrane, which we treat as an elastic medium since this
conforms with experimental findings in Dd \citep{Luo:2013:MMC}, we compute its
first variation with respect to a deformation, which gives the membrane force,
and to this we add the cortical forces directly.

The membrane has four modes of deformation: dilatation, shear, bending and
torsion, but only the bending mode is treated in general and we follow this
practice here. In addition to the bending energy, which to lowest order is
proportional to the square of the local curvature of the membrane, there are
contributions to the free energy corresponding to the work associated with area
and volume changes when these are conserved.  

We let $\Omega \subset  R^3$ denote the volume occupied by the cell and let $\cS$
denote its boundary. We assume that $\cS$ is a smooth, compact, two-dimensional manifold without boundary, parameterized by the map $\Phi:D \subset R^2 \rightarrow \cS$, defined so that the position vector $\bfx$ to any point on the membrane is given by $ \bfx = \bfx(u^1,u^2)$ for a coordinate pair $(u^1,u^2) \in D$.  We let $\bfn$ denote the outward normal on $\cS$, and define basis vectors on the surface via
\begin{equation}
\bfe_i = \dfrac{\partial \bfx }{\partial u^i} \qquad i = 1,2.
\label{basis}
\end{equation}
In general these are not normalized. 

The free energy associated with bending, which was first set forth for membranes
by \citet{Canham:1970:MEB} and later by \citet{Helfrich:1973:EPL}, has the
following form
\begin{eqnarray}
\label{freeng1} 
\cF_B=\int_{\cS} 2 k_c(H-C_0)^2 dS+\int_{\cS}k_G K dS.
\end{eqnarray} 
Here $\kappa_1$ and $\kappa_2 $ are the principal curvatures, $H= -(\kappa_1 +
\kappa_2)/2$ is the mean curvature, and $K = \kappa_1\kappa_2$ is the Gaussian
curvature\footnote{This definition of the mean curvature is predicated on
  choosing the outward normal as the normal to the surface.}. $C_0$ is a
phenomenological parameter called the spontaneous curvature, $k_c$ the bending
rigidity - which may be stress-dependent \citep{Diz-Munoz:2016:MTA}, and $k_G$
the Gaussian rigidity, which may also vary over the membrane. When $k_G$ is
constant, the integral of the Gaussian curvature is constant if $\Omega$ does
not change topological type under deformation, and the integral can be ignored.

 Under the constraints of constant surface area $A_0$ and
volume $V_0$ of the cell, the free energy takes the form
\begin{eqnarray}
\label{freeng2} 
\cF =  \cF_B  +
\int_D \Lambda \left(\sqrt{g} - \sqrt{g_0}\right) du^1du^2  + P \left(\int _{\Omega} dV  -V_0\right), 
\end{eqnarray} 
where $P \equiv P_{\textrm{ext}}-P_{\textrm{in}}$ is the pressure difference
across the membrane, which we assume is constant over the membrane. Typically
$P_{\textrm{in}}$ is a few hundred pascals higher than $P_{ext}$
\citep{Salbreux:2012:ACM}.  The constant term $P V_0$ simply translates the free
energy and can be ignored, since it disappears after the first variation of
(\ref{freeng2}) is taken. In the second-last integral over the boundary, $g$ is
the determinant of the metric tensor $\mathbf g$ of the surface, whose
components are $g_{ij} = \bfe_i \cdot \bfe_j$, and $g_0$ is its value in a
reference or undeformed configuration in which the area is $A_0$.
The integral represents the energy needed to alter the area from its initial
value, and $\Lambda$ is the energy per unit area that arises when $A$ is
changed.  Thus $\Lambda$ has units of force/length, which defines a tension, but
it is not a surface tension in the usual sense. Instead it is  an in-plane stress of
a two-dimensional surface, which  can be seen as follows. 

Consider a small x-y
section of a thin flat plate of thickness h, and suppose there are no normal
stresses to the section in the z-direction, which is orthogonal to the
plate. Suppose the plate is an elastic material and let $\stress$ be the stress
tensor and let $\bfeps$ be the strain relative to a reference configuration. For
small, uniform strains in the x-y plane on the four faces of the section one has
\begin{align}
\bfeps_x &= \dfrac{1}{E}(\stress_{xx} -\nu \stress_{yy}), \\
\bfeps_y &= \dfrac{1}{E}(\stress_{yy} -\nu \stress_{xx}),
\end{align}
where $E$ is Young's modulus and $\nu $ is the 
Poisson ratio of the material. From these it follows that 
\begin{align}
\stress_{xx} &= \dfrac{E}{1-\nu^2}(\bfeps_{x} + \nu \bfeps_{y}), \\
\stress_{yy} &= \dfrac{1}{1-\nu^2}(\nu \bfeps_{x} + \bfeps_{y} ).
\end{align}
If the stress and strains in the x and y directions are equal and uniform in the
z-direction, one can define a tension $T$ as 
$$
T = (\stress_{xx} + \stress_{yy})h = \dfrac{Eh}{1-\nu}\epsilon
$$
where $\epsilon = \bfeps_x + \bfeps_y.$ Experimental results are usually
reported in terms of a tension, but the foregoing shows that this assumes a
local isotropy of the stress and strain.  \cite{Evans:1987:PPS} define the
tension as the average of the stresses along the principal directions of the
surface, but these are rarely available.  It is shown later that in the absence
of imposed forces $\Lambda$ is constant, and therefore the constant area term
can be ignored as well.

In the absence of external forces, a stable equilibrium shape of a cell is a
minimizer of $\cF$, and thus a solution of $\delta^{(1)}\cF/\delta \bfx=0$ for
any infinitesimal deformation $\bfx = \bfx_0 + {\bphi}+ \psi \bfn$, where $
{\bphi} = \phi^i\bfe_i$,  of
$\cS$.  This leads to the following shape equations for the normal and tangential
components of the membrane force.\footnote{See Appendix A for a sketch of the derivation of these 
equations. When the bending and Gaussian moduli are constant the equation for
the normal deformation has been derived by \citet{Ouyang:1989:BEV,Capovilla:2003:DGL,Tu:2004:GTE} and others.}
\begin{eqnarray}
         F^{n} &=& -\frac{\delta \cF}{\delta\psi} = \big\{-\Delta_s\left[k_B\left(2H-C_0\right)\right]-k_B\left(2H-C_0\right)\left(2H^2+C_0H-2K\right) -\overline{\Delta_s}\, k_G+2\Lambda H-P\big\} \\
\vspace*{15pt} 
     F_i^{t} &=& -\frac{\delta \cF}{\delta \phi^i} =
     \bigg\{\frac{1}{2}\left(2H-C_0\right)^2\nabla_i k_B+K\nabla_i k_G+\nabla_i\Lambda\bigg\}
     \qquad  i = 1,2.
 \label{shape1}
\end{eqnarray}
%

Here $\Delta_s$ and $\bar{\Delta}_s$ are surface Laplacians ({\em cf.~}Appendix A),
and the $\nabla_i$ are the  components of the
surface gradient, resp. In the first equation one sees that $\Lambda$ enters the
normal component via the term $2\Lambda H$, which couples areal distension to
the curvature in the normal component of the force. In light of how the
variation is defined, the resultant forces are defined per unit area.

To simplify the equations we assume hereafter that $C_0 = 0$, which reduces the foregoing to 
\begin{eqnarray}
          \label{shape20}
         F^{n} &=& -\frac{\delta \cF}{\delta\psi} =-2\Delta_s(k_BH)-2k_BH\left(H^2-K\right) -\overline{\Delta_s}\, k_G+2\Lambda H-P \\
\vspace*{15pt} 
     F_i^{t} &=& -\frac{\delta \cF}{\delta \phi^i} =
     2H^2\nabla_{i}k_B+K\nabla_{i} k_G+\nabla_{i}\Lambda
     \qquad  i = 1,2.
 \label{shape2}
\end{eqnarray}
When the bending and Gaussian rigidities are constant
these  simplify to 
\begin{eqnarray}
         F^{n} &=& -\frac{\delta \cF}{\delta\psi} =-2k_B\Delta_sH
         -2k_BH\left(H^2 - K\right) +2\Lambda H-P \\
\vspace*{15pt} 
     F_i^{t} &=& \nabla_{i}\Lambda
     \qquad  i = 1,2.
 \label{shape3}
\end{eqnarray}
In any case the forces vanish at critical points of the energy, and stable shapes
correspond to  local minima of the  free energy.

When written in $(u^1,u^2)$ coordinates, the first term in $F^n$ defines a
fourth-order differential operator, and the critical points of the energy cannot
in general be found analytically. Since the membrane is embedded in a quiescent
viscous fluid, we define a fictitious relaxation process for the evolution of
the shape from its initial values in which we suppose that the dominant force is
viscous drag, and we neglect inertial effects. This leads to the system
\begin{eqnarray}
{\mu}_d  \dfrac{d \psi (u^1,u^2)}{d\tau}& =& F^n(H,K,P, \Lambda,k_B,k_G,u^1,u^2) \\
\label{evo1}
{\mu}_d  \dfrac{d \phi^i(u^1,u^2)}{d\tau} &=&   F_i^t(H,K,P, \Lambda,k_B,k_G,u^1,u^2)  \qquad i = 1,2
\label{evo2}
\end{eqnarray}
where  ${\mu}_d$ can be thought of as a drag coefficient with the dimension of the viscosity per unit length. 

\subsection{The shape equations under  cortical forces} 
\label{shapecort}

To incorporate cortical forces, which are non-conservative, we add these to
the intrinsic membrane forces as prescribed normal and
tangential forces per unit area, the components of which are denoted $f^n$ and $f^t_i, i
=1,2$. Thus the equations to be solved for a 3D shape are
\begin{eqnarray}
\label{evo3}
 {\mu}_d \dfrac{d \psi (u^1,u^2)}{d\tau} &=& F^n(H,K,P, \Lambda,k_B,k_G,u^1,u^2)  + f^n\\
 {\mu}_d \dfrac{d \phi^i(u^1,u^2)}{d\tau} &=& F_i^t(H,K,P, \Lambda,k_B,k_G,u^1,u^2) +f_i^t  \qquad i = 1,2.
\label{evo4}
\end{eqnarray}
Since the normal force is directed inward and the  normal vector is directed
outward,  $f^n < 0$.  When the cortical forces are incorporated the resulting evolution is no longer a
gradient flow, and one simply looks for steady states of (\ref{evo3}) and
(\ref{evo4}), which in general are not minimizers of the membrane free energy.

The equations can be cast into non-dimensional form by defining the * variables
$~H^* = HR_0, ~L^*= L/R_0, ~\Delta^* = R_0^2\Delta, ~k_B^* = k_B/ 
\bar{k}, ~k_G^* = k_G/\bar{k}, ~\Lambda^*= R_0^2 \Lambda/\bar{k}, ~P^* = R_0^3
P/\bar{k}$, $~f^{n*} = f^n R_0^3/\bar{k}~$, $~f_i^{t*} = f_i^t R_0^3/\bar{k}~$
and $~{\tau}^*= \tau/\tau_{\rm m}$, where $R_0$ and $\bar{k}$ are
the characteristic length and energy, resp., which we set equal to 1. Therefore,  $\tau_{\rm
    m} \equiv \mu R_0^4/\bar{k}$ defines a characteristic time unit scaled by
a constant characteristic bending rigidity unit $\bar{k}$. The resulting forms of (\ref{evo3})
and (\ref{evo4}) in unstarred variables  are then
\begin{eqnarray}
\label{evo5}
\dfrac{d \psi (u^1,u^2)}{d\tau}& =& F^n(H,K,C_0,P, \Lambda,k_B,k_G,u^1,u^2)  + f^n\\
\dfrac{d \phi^i(u^1,u^2)}{d\tau} &=&   F_i^t(H,K,C_0,P, \Lambda,k_B,k_G,u^1,u^2) +f_i^t  \qquad i = 1,2.
\label{evo6}
\end{eqnarray}

\section{ 2D shapes }

\subsection{The evolution equations for 2D shapes }

 In order to understand the effects of membrane  and cortical forces in
 determining the shape in the simplest possible context, we first perform an
 analysis of the cell shapes in two dimensions. In this case the  domain
 is a 2D area, the  boundary $\cS$ is a  closed curve, and the area and volume
 constraints become perimeter 
 and area constraints. The energy functional now reads 
\begin{eqnarray}
\label{freeng2d} 
\cF^{2D}=\int_{\cS} \frac{k_B}{2}\kappa^2 ds + \int_{\cS} \Lambda \left(\sqrt{g} - \sqrt{g_0}\right) ds  + P \left(\int _{\Omega} dA  -A_0\right), 
\end{eqnarray} 
where $\kappa$ is the curvature and $s$ is the arc length on the boundary.  The normal
and tangential forces  arising from the bending energy and constraints  reduce
to 
\begin{eqnarray}
  \label{2dshape1}
     F^{n} &=& -\frac{\delta \cF^{2D}_B}{\delta\psi} =-\Delta_s(k_B\kappa)-\frac{k_B}{2}\kappa^3 + \Lambda \kappa-P \\
\vspace*{15pt} 
     F^{t} &=& -\frac{\delta \cF^{2D}_B}{\delta \phi^i} =
     \frac{\kappa^2}{2}\nabla k_B+\nabla \Lambda,
 \label{2dshape2}
\end{eqnarray}
and the evolution equations
 (\ref{evo5}) and (\ref{evo6}) now read 
\begin{eqnarray}
\label{2devo5}
\dfrac{d \psi (s)}{d\tau} &=& F^{n}+f^n= -\Delta_s(k_B\kappa)-\frac{k_B}{2}\kappa^3 + \Lambda \kappa-P + f^n\\
\dfrac{d \phi(s)}{d\tau} &=& F^{t}+f^t= \frac{\kappa^2}{2}\nabla k_B+\nabla \Lambda +f^t .
\label{2devo6}
\end{eqnarray}

\subsection{The numerical algorithm for solving the evolution equations}
\label{QD2CS}

As a characteristic length scale of
the problem we use an effective 2D cell size $R_0$, which is defined via 
\begin{eqnarray}
A=\pi R_0^2
\label{area}
\end{eqnarray}
where $A$ is the specified area enclosed by the cell contour, which has total
fixed length $L$. The cell shape can then be characterized by the reduced area
\begin{eqnarray}
\Gamma=\frac{A}{\pi\left(\frac{L}{2\pi}\right)^2}=\frac{4\pi A}{L^2},
\label{reduced}
\end{eqnarray}
which is an intrinsic dimensionless parameter that expresses the ratio of the
area of a given 2D shape to that of a circle of circumference $L$ and a
larger area. 

To solve the 2D eqns (\ref{2devo5}) and (\ref{2devo6}), the boundary curve is
discretized into $N$ segments \citep{Wu:2009:TNI} and we assume a 'time' step,
$\tau_0$ and label the time sequence as
$\tau_0,2\tau_0,\dots, j\tau_0$. The vectorial
form of eqns (\ref{2devo5}) and (\ref{2devo6}) can be concisely transformed into
\begin{equation}
\mathbf{r}_{j+1}^k-\mathbf{r}_{j}^k=\mathbf{f}_j^k \tau_0,
\label{disrelaxeq}
\end{equation}
where $\mathbf{f}_j^k$ is the discrete form of the dimensionless force surface
density. The geometric quantities in this equation can be discretized as:
\begin{eqnarray}
[ds]_j^k&=&|\mathbf{r}_{j}^{k+1}-\mathbf{r}_{j}^k|,
\quad [\mathbf{t}]_j^k=\frac{\mathbf{r}_{j}^{k+1}-\mathbf{r}_{j}^k}{[ds]_j^k}
\quad [\mathbf{n}]_j^k= - \frac{\mathbf{t}_{j}^{k+1}-\mathbf{t}_{j}^k}{[\kappa\ ds]_j^k},\\ [5pt]
\label{localcurv}
[\kappa]_j^k &=& \frac{\sqrt{ (x_{j}^{k+1}-2x_{j}^k+x_{j}^{k-1})^2 + (y_{j}^{k+1}-2y_{j}^k+y_{j}^{k-1})^2} }{\left\{[ds]_j^k\right\}^2}.
\end{eqnarray}
\begin{eqnarray}
\label{curvforce}
[\Delta_s\kappa]_j^k &=& \frac{\kappa_{j}^{k+1}-2\kappa_{j}^k+\kappa_{j}^{k-1}}{\left\{[ds]_j^k\right\}^2}.
\end{eqnarray}
where $\mathbf{r}_j^k=x_j^k\mathbf{e}_x+y_j^k\mathbf{e}_y$ and $\mathbf{n}_j^k$ can also be obtained by rotating $\mathbf{t}_j^k$ in $90^\circ$
counterclockwise. 

The local invariance of the boundary length  is treated using a spring-like penalty method
instead of introducing a spatially variable Lagrange multiplier, but our method
is equivalent to it, as shown in Appendix B. One can re-define the areal  energy
density via an harmonic spring-like potential, which leads to a simple
implementation. Let  
$$
\Lambda =\Lambda_0 \left(\sqrt{\frac{g}{g_0}}-1\right), 
$$
where $\Lambda_0$ is a constant penalty coefficient. Then 
\begin{equation}
\mathbf{f}^{\rm T}(s) = \frac{\partial}{\partial s}\left[\Lambda_0\left(\sqrt{\frac{g}{g_0}}-1\right)\mathbf{t}(s)\right],
\label{2dten}
\end{equation}
which can be rewritten as
\begin{equation}
\mathbf{f}^{\rm T}(s) = \frac{\Lambda_0}{ds_0}\frac{\partial}{\partial s}\left[\left(ds-ds_0\right)\mathbf{t}(s)\right].
\label{penalten2}
\end{equation}
 Eq(\ref{penalten2}) can be discretized as
\begin{equation}
\mathbf{f}_j^{{\rm T},k} = \frac{N\Lambda_0}{L}\frac{[\left(ds-ds_0\right)\mathbf{t}]_{j}^{k+1}-[\left(ds-ds_0\right)\mathbf{t}]_{j}^k}{[ds]_j^k},
\label{numerpenal}
\end{equation}
where $ds_0=L/N$ and $L$ is the total contour length of the  2D cell.

Because $\Lambda_0$  can be considered as a membrane tension, one can
define a new time scale ${\tau}_{\rm T}=\mu_{\rm o} R_0/\Lambda_0$. This is to be
compared to the membrane bending time scale ${\tau}_{\rm m}$ defined earlier,
and must be taken small enough in comparison to ${\tau}_{\rm m}$ to ensure
invariance of the boundary length on the membrane time scale. For most practical
purposes ${\tau}_{\rm T}=10^{-5}-10^{-4}{\tau}_{\rm m}$ has proven to be
sufficiently small \citep{Wu:2015:AMC,Wu:2016:ASC}.

One could treat $\Lambda_0$ as a true Lagrange multiplier, and develop a scheme
to simultaneously solve for the shape and multipliers, but we simply fix a
large $\Lambda_0$ and check {\em en passant} that the arc length is conserved
to the desired precision. The  area constraint is treated as described in
Appendix C.  The stopping condition of the time-stepping  calculation in (\ref{disrelaxeq})
is $\sum_{k=1}^N|\mathbf{r}_{J+1}^k-\mathbf{r}_{J}^k|<\epsilon$ for a large
enough integer $J$ and a small enough number $\epsilon$ ($\epsilon=10^{-6}$ is
used here).

A summary of the computational procedure is as follows. First, the force
distribution along the cell membrane is computed based on the current
configuration of the cell. Second, the evolution equations (\ref{2devo5}) and
(\ref{2devo6}) are used to obtain the new position  in terms of the
normal and tangential force distributions.  Third, the first two steps are
repeated until the stopping criterion has been satisfied, indicating that the
steady state has been reached.

\subsection{Computational results for 2D shapes}

\begin{figure}[h!]
\hspace*{10pt}
\includegraphics[width=5.25in]{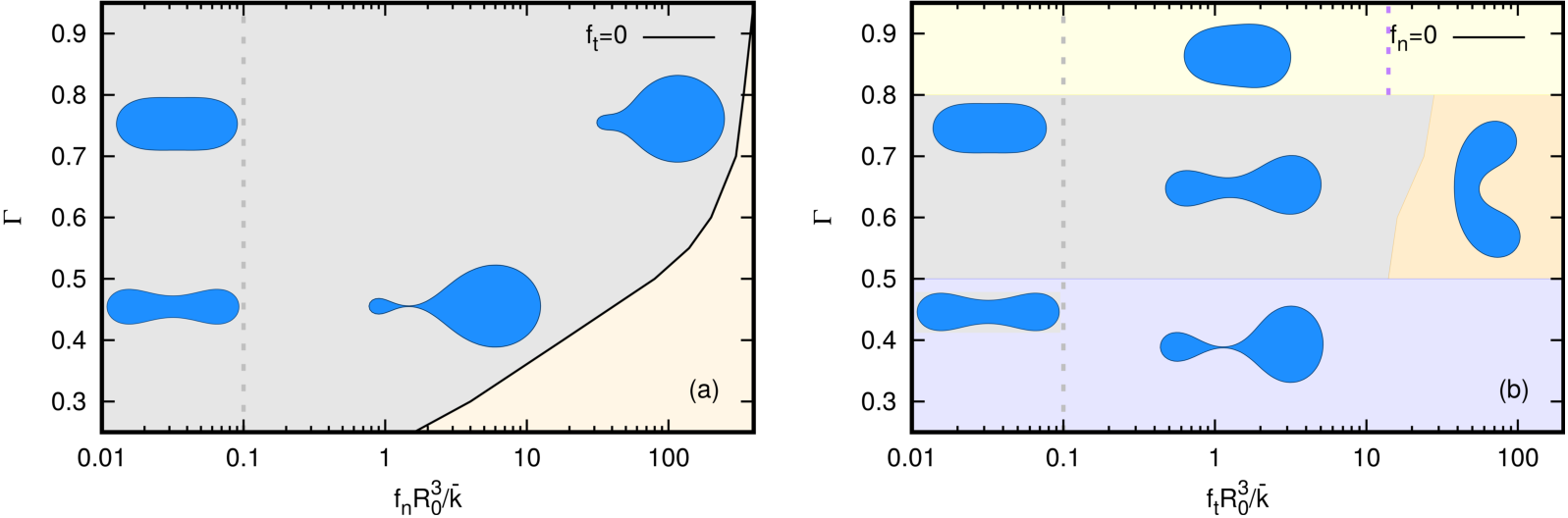}
\caption{ (a) A phase diagram showing cell shapes as a function of the
 dimensionless normal force $f^*_n$ and the reduced area $\Gamma$. (b) A
  similar diagram for the dimensionless   tangential force $f^*_t$ and the
  reduced area $\Gamma$. The dimensionless range $0.1 \sim 100$ in the ordinate
  corresponds to dimensional forces in the range $0.01 \sim 10$ nN/$\mu m$ for
  the chosen cell size and bending modulus.}
\label{2Dshape}
\vspace*{-10pt}
\end{figure}

In this section we investigate the effects of spatial variations in the imposed
cortical forces and in the bending modulus.  Figure \ref{2Dshape}(a) shows
representative shapes as a function of the magnitude of the dimensionless tether
force $f_n^*$ and the reduced area $\Gamma$, while Figure \ref{2Dshape}(b) shows
the shapes as a function of the dimensionless tangential force $f_t^*$ and
$\Gamma$. In this figure and the following one, the property in question varies linearly from right to left of the shapes, with a minimum of zero at the right and the maximum given by the value on the left in the x-axis. As a point of reference, if $\Gamma = 1$ all shapes are disks, irrespective of the variation of other
properties. In both panels one sees that there is little effect on the shape
when either force is less than ~0.1, but a very strong effect for either force
greater than 1. Shapes in the gray zone on the left are stable, but no stable
shapes were found in the light yellow zone due to numerical difficulties. Stable
shapes are found for the entire range of the tangential force on the right.

The normal force (a) leads to pear-shaped cells at large $\Gamma$ and shapes
with two lobes connected by a narrow 'bridge' at $\Gamma \sim 0.45$, the latter somewhat similar to the Ruprecht 'stable-bleb' type shown in Figure \ref{2morph}(a).  This shape only exists at a sufficiently small reduced area, when the shape under a small force is biconcave, and stems from the large normal force at the right end of the cell. A variety of other shapes can be obtained under different variations of the tether force. In particular, a symmetric two-lobed shape can be obtained by concentrating the normal force force at the center of the cell.

At $f_t^* \sim 1$ and small and intermediate $\Gamma$ in the panel on the right,
one sees that the large tangential component at the left end leads to a smaller lobe there, while the small tangential force at the right end leads to a larger lobe. These are again similar to the stable-bleb type observed experimentally,
where it is suggested that the cortical tension is lowest at the front of the cell (the right end in the figure) and highest at the rear. Clearly the curvature
is highest on the small lobe, which can be understood from the fact that a high
tangential tension enters into the normal component of the membrane force
equations. While the stable shapes are not necessarily minimizers of the area,
the membrane-derived component of the forces drives the evolution toward a
minimum.  The tension energy is highest at the right end of the cell while the
bending energy dominates the rear. Since the tension is lowest at the right end,
the evolution there is driven by a tendency toward local energy minimization,
which leads to larger radii to minimize the local curvature.  If the magnitude
of the tension increases further, the pear-like shape becomes unstable and
evolves into a kidney shape with a shallow indentation or a kidney shape with a
deep one, the latter shown in the light orange zone, depending on the applied
magnitude and the reduced area. The kidney shape cannot be attained at low
$\Gamma$ due to the narrow neck that occurs around $f^*_t \sim 1$, which
precludes the shape changes involved in the transition to the kidney shape. The kidney shapes have been extensively found in the biophysical experiments of giant vesicles \citep{Lipowsky2014} as well as some chemically treated erythrocytes, such as cupped red cells \citep{Brailsford1980}.
\begin{wrapfigure}[]{r}{2.5in}
\includegraphics[width=2.5in]{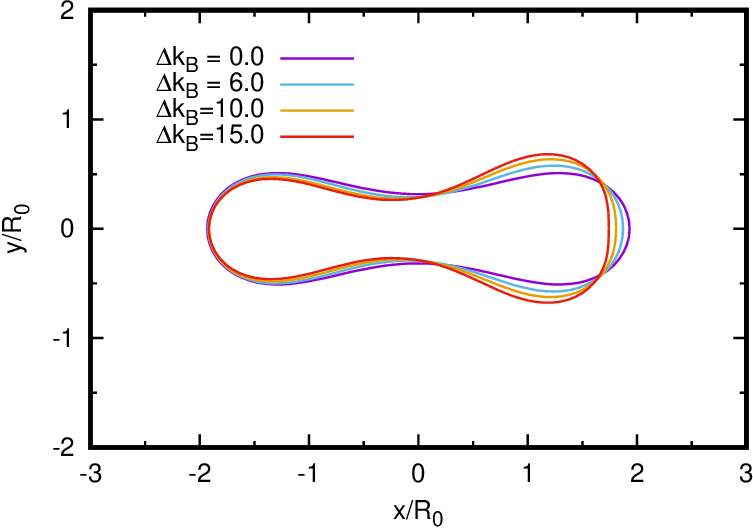}                                
\caption{  The dependence of shape on the magnitude of the 
variation in $k_B$.  The shape changes  from 
    biconcave to the pear shape as the axial variable bending rigidity increases.  $\Gamma=0.6$ } 
\label{2Dshape3}
\vspace*{-20pt}
\end{wrapfigure}

Experimental measurements of cortical tensions fall in a wide range, from 0.02
nN/$\mu$m for neutrophils to 4.1 nN/$\mu$m for Dd
\citep{Winklbauer:2015:CAS}. Estimates based on the number of linker proteins
give a tether force of about 1 nN/$\mu$m \citep{Diz-Munoz:2010:CDC}. Our model predicts a reasonable experimental range 0.01 $\sim$ 10 nN/$\mu$m
for the chosen cell size $R_0$ = 10$\mu$m. The results in Fig. 5 are given in dimensionless terms, and a comparison of them with the experimental results shows an excellent agreement. 

To investigate the effect of a variable bending modulus, we vary it 
axially using the hyperbolic tangent function
\begin{eqnarray}
k_B=\bar{k}+\frac{\Delta k_B}{2}\left[1+\tanh\left(\frac{s}{w}\right)\right],
\label{varbending}
\end{eqnarray}
where $s$ is the  length of the half contour measured from the rightmost end of the cell, $\bar{k}$
is the uniform component of the bending energy,  $\Delta k_B$ is the bending
rigidity difference between the right  and left ends of the cell, and $w$ is the
width of the transition zone. In a thin-shell description of a material the
Young's modulus varies as the thickness cubed, and hence the axial variation
imposed here could arise from the axial variation of the cortical thickness.  It
might also reflect a distribution of adsorbed or transmembrane proteins, and of
glycolipids \citep{Wu:2013:MPM,Wu2013effects}.  Figure \ref{2Dshape3} shows
that in a quiescent fluid  a cell undergoes a
shape transition from the biconcave shape to the pear shape induced by the
variable bending rigidity. Since there are no imposed forces here the shapes are
minimizers of the free energy. One sees that the cell 'expands' in regions of
large rigidity since a region of  higher $k_B$  offsets the lower curvature in 
the free energy $\cF$.

\section{Tension-driven swimming at low Reynolds number}
\label{Sec.MD}

\subsection{The boundary integral equation}

When the cell is submerged in a fluid and free to move, it generates extra- and
intracellular flows, and here we assume that both the extra- and intracellular
fluids are Newtonian, of density $\rho$ and viscosity $\mu$. In reality both
fluids are undoubtedly more complex, but to demonstrate that movement ensues
under tension gradients in the membrane it suffices to consider Newtonian
fluids.  

The Navier-Stokes equations for the intra- and extracellular fluid
velocities $\bfu$ are \citep{Childress:1981:MSF}
\begin{align}
\label{NSeqn}
\rho\dfrac{\partial \bfu}{\partial t} +\rho (\bfu \cdot\nabla) \bfu &=  \nabla
 \cdot  \sig  + \bff_{\textrm{ext}}
=- \nabla p  + \mu \Delta \bfu + \bff_{\textrm{ext}} , \\
   \nabla \cdot \bfu &= 0
\end{align}
where 
$$
\sig = -p \bm{\delta} + \mu (\nabla \bfu + (\nabla \bfu)^T)
$$ 
is the Cauchy stress tensor and ${\bf f}_{\textrm{ext}} $ is the external force
field. We assume that the intra- and extracellular densities are equal, and thus the
cell is neutrally buoyant, but allow for different viscosities in the two
fluids.  We further assume that there are no additional body forces and thus set
$\bff_{\textrm{ext}} = 0$.

When converted to dimensionless form and the symbols re-defined, these
equations read
\begin{eqnarray}
\label{Eq.ReSl_Stokes}
\textrm{ReSl} \dfrac{\partial \bfu}{\partial t} + \textrm{Re} (\bfu \cdot\nabla) = - \nabla p + \Delta \bfu, 
\quad \quad \nabla \cdot \bfu = 0,
\end{eqnarray}
where the Reynolds number Re based on a characteristic length scale $L$ and a
characterisic speed scale $U$ is Re = $\rho$LU /$\mu$.  In addition,
$\textrm{Sl} = \omega L/U$ is the Strouhal number and $\omega$ is a
characteristic rate at which the shape changes. When $\textrm{Re} \ll 1$ the
convective momentum term in equation (\ref{Eq.ReSl_Stokes}) can be neglected,
but the time variation requires that $\textrm{ReSl} = \omega \rho L^2/\mu \ll 1$,
which implies that the initial shape changes must be slow enough.  When both
terms are neglected, which we assume throughout, a low Reynolds number (LRN)
flow is governed by the Stokes equations, now in dimensional form, 
\vspace*{-5pt}
\begin{equation} 
 \mu \Delta \bfu - \nabla p = {\bf 0}, \qquad \qquad \nabla
\cdot \bfu = 0.
\label{creep}
\end{equation}

Here we consider small cells such as Dd, whose small size and low speeds lead to
LRN flows \citep{Wang:2015:CAA}, and in this regime time does not appear
explicitly, there are no inertial effects, and bodies move by exploiting the
viscous resistance of the fluid.  In the absence of external forces on the fluid
and acceleration of the swimmer, there is no net force or torque on a
self-propelled swimmer in the Stokes regime, and therefore movement is a purely
geometric process.

The solution of the Stokes equations via the boundary integral method is a
well-studied problem for red blood cells or vesicles in a pressure-driven flow
\citep{Zhao:2010:SBI,Veerapaneni:2011:FAS,Rahimian:2015:BIM}, but here the
tension that arises from the active contraction of the cortex drives the flow.
In the boundary integral method (BIM) one uses an Eulerian description for the
velocity field of the fluid and a Lagrangian description for the configuration
of the membrane. We assume that the interior and exterior viscosities are equal
and that  the velocity at infinity vanishes, and therefore the solution
has the representation \citep{Pozrikidis:1992:BIS,Pozrikidis:2003:MSC}
\begin{eqnarray}
\mathbf{u}(\bfx) =  \dfrac{1}{4\pi\mu}\int_{\cS(t)}
G(\bfx,\bfx_0) \cdot \bF_m(\bfx_0)  ds
\label{stsol}
\end{eqnarray}
where 
\begin{eqnarray}
G_{ij}\left(\mathbf{x},\mathbf{x}_0\right)=-\delta_{ij}\ln |\mathbf{x}-\mathbf{x}_0|+\frac{\left(\mathbf{x}-\mathbf{x}_0\right)_i \left(\mathbf{x}-\mathbf{x}_0\right)_j}{|\mathbf{x}-\mathbf{x}_0|^2}
\label{2dG}
\end{eqnarray}
is the Green's function in two space dimensions and $\bF_m$ is the force exerted
by the membrane on the fluid. We assume continuity of the velocities across the
membrane, and mechanical equilibrium at the membrane, and therefore the force
balance reads
\vspace*{-5pt}
\begin{equation}
\bF_m \equiv \big [ (\sig_{in}  - \sig_{ext}) \cdot \bfn \big ]_m =\bff_m.
\label{forbal}
\end{equation}
 The right-hand side of (\ref{forbal}) is the
force on the membrane given by eqns (\ref{2devo5}) and (\ref{2devo6}), and thus one
has to solve the integral equation, the shape equation, and the tangential force
balance to determine the interior,  exterior and boundary velocity fields.
 
When $\bfx \to \bfx_0 \in \cS (t)$, the Green's function  in (\ref{2dG})
diverges, but it is a weak singularity, and can be removed by using the
singularity subtraction transformation (SST) \citep{Farutin:2014:NSV}, which is
based on the integral identities
\begin{eqnarray}
& \mathbf{n}(\mathbf{x})\cdot\int_{\cS}
G(\mathbf{x},\mathbf{x}_0)\cdot\mathbf{n}(\mathbf{x}_0)
ds=0, \\ [5pt]
& \mathbf{t}(\mathbf{x})\cdot\int_{\cS} G(\mathbf{x},\mathbf{x}_0)\cdot\mathbf{t}(\mathbf{x}_0) ds+\dfrac{\mathbf{n}(\mathbf{x})}{2\pi}\int_{\cS} \dfrac{(\mathbf{x}-\mathbf{x}_0)\left[(\mathbf{x}-\mathbf{x}_0)\cdot\mathbf{n}(\mathbf{x}_0)\right]}{|\mathbf{x}-\mathbf{x}_0|^2} ds=0
\label{identity}
\end{eqnarray}
where $\mathbf{t}$ is the tangent vector to the contour ${\cS}$. The first
identity follows from the divergence theorem and the second from Stokes'
theorem. The SST technique for the single-layer kernel
$G(\mathbf{x},\mathbf{x}_0)$ can be applied as follows.  Rewrite the integral in
(\ref{2dG}) as
\begin{eqnarray}
&\int_{\cS}& G(\mathbf{x},\mathbf{x}_0) \cdot \mathbf{f}(\mathbf{x}_0) ds  \nonumber 
= \\  [7pt]
&\int_{\cS} &[ G(\mathbf{x},\mathbf{x}_0) \cdot
\mathbf{f}(\mathbf{x}_0) -
\mathbf{f}_n(\mathbf{x}_0)\mathbf{n}(\mathbf{x}) \cdot
G(\mathbf{x},\mathbf{x}_0)\cdot\mathbf{n}(\mathbf{x}_0) \nonumber
- \mathbf{f}_t(\mathbf{x}_0)\mathbf{t}(\mathbf{x}) \cdot G(\mathbf{x},\mathbf{x}_0)\cdot\mathbf{t}(\mathbf{x}_0)] ds \nonumber \\[7pt]
&-& \frac{\mathbf{f}_t(\mathbf{x})}{2\pi}\int_{\cS}\frac{\left[(\mathbf{x}-\mathbf{x}_0)\cdot\mathbf{n}(\mathbf{x})\right]\left[(\mathbf{x}-\mathbf{x}_0)\cdot\mathbf{n}(\mathbf{x}_0)\right]}{|\mathbf{x}-\mathbf{x}_0|^2} ds,
\label{SST}
\end{eqnarray}
where
$\mathbf{f}(\mathbf{x}_0)=f_n(\mathbf{x}_0)\mathbf{n}(\mathbf{x}_0)+f_t(\mathbf{x}_0)\mathbf{t}(\mathbf{x}_0)$
and
$\mathbf{f}_n(\mathbf{x}_0)=(\mathbf{f}(\mathbf{x}_0)\cdot\mathbf{n}(\mathbf{x}_0))\mathbf{n}(\mathbf{x}_0)$
and
$\mathbf{f}_t(\mathbf{x}_0)=(\mathbf{f}(\mathbf{x}_0)\cdot\mathbf{t}(\mathbf{x}_0))\mathbf{t}(\mathbf{x}_0)$
are the normal and tangential parts of the force
$\mathbf{f}(\mathbf{x}_0)$. Both integrals on the RHS of (\ref{SST}) have no
singularities, that is they both go to zero when $\mathbf{x}$ approaches
$\mathbf{x}_0$ along the contour ${\cS}$. The second term in (\ref{SST}) tends
to zero  when $\mathbf{x}_0$ approaches $\mathbf{x}$ along the contour because
$(\mathbf{x}_0-\mathbf{x})\cdot\mathbf{n}\approx\cO(|\mathbf{x}_0-\mathbf{x}|^2)$.

\subsection{The numerical algorithm for cell swimming}

 Step 1: Following the dimensionless quantities denoted with a star symbol in
 Section \ref{shapecort}, the dimensionless surface force   in unstarred
 variables $\mathbf{f}\left(\mathbf{x}_0\right)$ can be expressed in the
 component as the right hand side (RHS) of eqns (\ref{2devo5}) and
 (\ref{2devo6}) in terms of the current configuration of the cell.  \vspace*{5pt}

Step 2: By substituting the instantaneous surface force distribution into the dimensionless boundary integral  in unstarred variables along the cell shape contour, we can obtain the velocity of each material node on the membrane 
\begin{eqnarray}
\mathbf{u}\left(\mathbf{x}\right)=\frac{1}{4\pi}\int_{\cS} G\left(\mathbf{x},\mathbf{x}_0\right)\cdot\mathbf{f}\left(\mathbf{x}_0\right){\rm d}s\left(\mathbf{x}_0\right)
\label{bierewrite}
\end{eqnarray}
Step 3: The motion of cell membrane is obtained by solving the dimensionless time evolution equation  in unstarred variables
\begin{eqnarray}
\frac{d\mathbf{x}}{d\tau} = \mathbf{u}\left(\mathbf{x}\right)
\label{vel}
\end{eqnarray} 
for each material node $\mathbf{x}$ lying on the membrane and the new cell configration is updated at each time step using an explicit Euler scheme: 
\begin{eqnarray}
\mathbf{x}(t+{\rm d}\tau) = \mathbf{x}(\tau)+\mathbf{u}\left(\mathbf{x}(\tau),\tau\right){\rm d}\tau.
\label{displace}
\end{eqnarray}

Step 4: Repeat the former three steps until the stopping criterion has been
satisfied, indicating that the steady shape and swimming velocity have been
obtained.

The discretization procedure and validation of the method can be found in Appendix \ref{Appendix D}. 

\subsection{Computational results for cell swimming}

To compute the velocity field in the interior and exterior fluids we introduce a
square Eulerian grid with a specified degree of refinement -- the mesh size can
be taken significantly smaller than ${\rm d} s$ according to the accuracy
required. Since the Green's function is singular when the target point coincides
with the source point, a small strip (of order ${\rm d} s$ around the boundary
is excluded from computation of the velocity field. The Eulerian lattice grid points do not coincide with the Lagrangian nodes 
\begin{wrapfigure}[]{r}{2.5in}
   \includegraphics[width=2.25in]{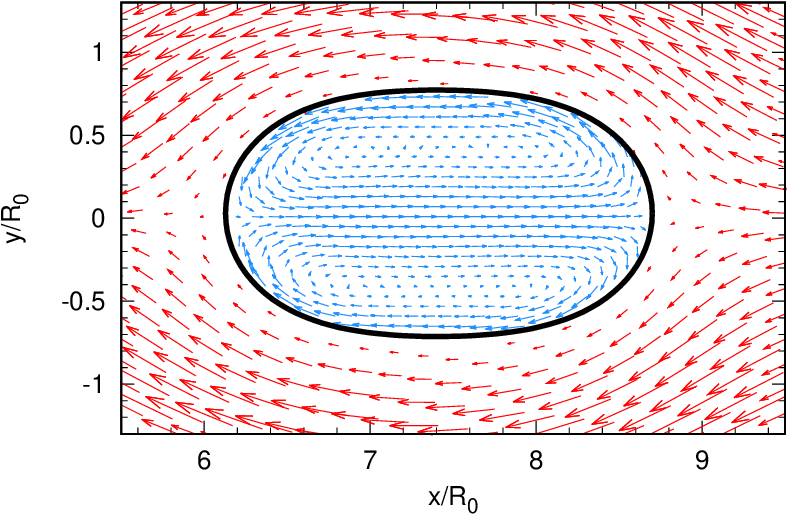} 
\caption{  The velocity field  inside and outside the swimming cell in
  cell-based coordinates.
    } 
\label{2Dvelc}
\vspace*{-20pt}
\end{wrapfigure}
on the membrane in general,
and we need only to evaluate the distance between the source point (lying on the
membrane) and the target point (lying on the square lattice grid). For each
point $\mathbf{x}$, the velocity field is evaluated by using (\ref{bierewrite}),
where the integral along the membrane is performed exactly. A simple but
accurate algorithm to judge the cellular interior and exterior is given in
Appendix \ref{Appendix E}.

\begin{figure}[h!]
   \includegraphics[width=5.in]{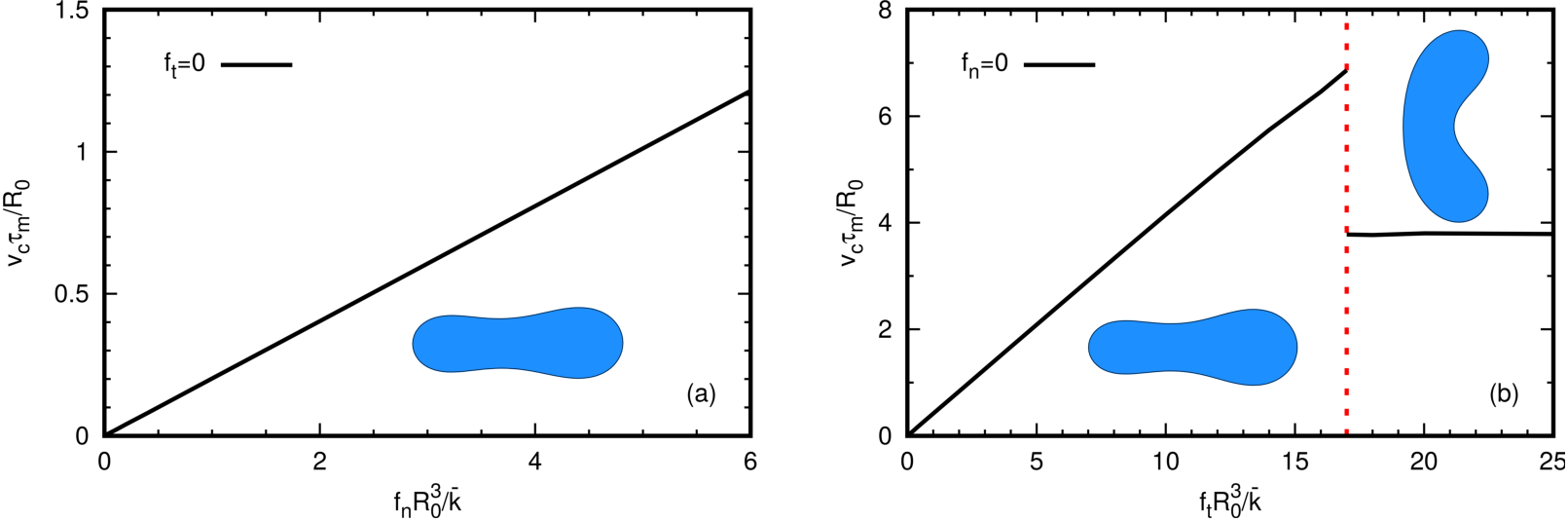} 
   \caption{ The swimming velocity as a function of the applied forces.    The gradient of
   the forces is reversed from earlier figures, and the  cells move to the
   right. In the both diagrams the cell has a reduced area $\Gamma=0.6$.}
\label{2Dvel}
\vspace*{-10pt}
\end{figure}

Figure \ref{2Dvelc} shows the velocity field outside the swimming cell in a
cell-fixed frame. The interior flow (not shown) indicates that tension driven
swimming can form a microcirculation flow inside the cell, which is consistent
with the biological conjecture shown in Figure \ref{cortflow}.  Figure
\ref{2Dvel}(a) shows that the swimming velocity as a function of the applied
forces is linear for all normal forces (a), and linear for small tangential
forces (b).  For large tangential forces, the speed drops abruptly in the
transition to a kidney shape.  The plateau shown in (b) is not precisely flat,
but the values are close.  
  
%

\begin{figure}[h!]
  \centerline{ \includegraphics[width=2.5in]{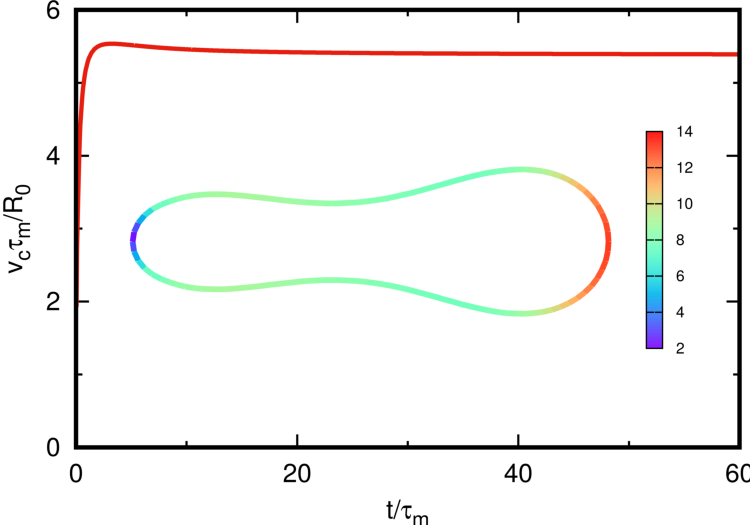} }
   \caption{The velocity, shape and surface force distribution on the cell. The
     cell shape has a reduced area $\Gamma=0.5$ with $f_t=0.0$, $f_n=0.0$, and
     $\Delta k_B=6.0$.  The cell moves in the direction of the large lobe. }
\label{2Dvelb}
\end{figure}

Figure \ref{2Dvelb} shows the velocity and evolution of cell shape as a function
of time. Initially the shape is biconcave, but evolves to the pear shape induced
by the variable bending rigidity, and the corresponding swimming velocity
changes with time. As in Figure \ref{2Dshape3}, the lobe size depends on the
bending rigidity difference between the front and the rear of the cell -- the
larger the difference, the larger the difference in lobe size. The inset shows
the force distribution on the boundary, where one sees that stress concentrates
on the large lobe. \hspace*{6in}

\section{Discussion}

Metastasis of a cancerous growth, which involves development of secondary
malignant growth distinct from the primary tumor, accounts for the poor
prognosis in many cancer types, and thus understanding how cells move in various
environments is an important step toward devising medical treatment strategies
for inhibiting metastasis. Because many cell types can use different modes
ranging from crawling to swimming, it is important to understand how the
micro-environment a cell finds itself in influences the mode it chooses.  The
shape a cell adopts in different environments is dependent on both intra- and
extracellular forces on the membrane, and the balance between them can in turn
determine the mode of movement via feedback of mechanical forces on the
structure of the cytoskeleton and cortex.  Earlier we described how certain cell types
generate cortical flows that can produce motion, but there is little
understanding of the quantitative relationships between the various forces, 
cell shape and movement.

Recently \citet{Callan-Jones:2016:CFS} addressed the question of how cortical
flows affect the shape of cells analytically, via a perturbation analysis of a
spherical cell. They find that for small perturbations of a sphere the cortical
force is the primary determinant of the shape, and showed that the flow can lead
to axial asymmetries of the shape, similar to those observed.  However a
complete treatment of the control of cell shape and movement is currently beyond
reach, either analytically or computationally, and in this paper we have
addressed several simpler questions, which are (i) how do the shapes adopted by
cells constrained in a quiescent fluid depend on prescribed cortical forces, and
(ii) how does a cortical tension gradient affect the shape and speed
of movement when it is unconstrained.
 
A primary objective of this work was to demonstrate that cells could swim
without any  shape changes, simply by using tension gradients in the
membrane. To show this, we developed a two dimensional mechanical theory that
captures the necessary cortex-membrane and fluid-membrane interactions. This
enabled us to analyze the range of steady state shapes that could be adopted by
membranes under tension distributions and whether these distributions could lead
to swimming via an interaction with the fluid. We found that the resulting
shapes closely resemble some of the shapes observed experimentally.  Several new
features of cell deformation and microswimming have emerged, including the
following.  (1) The pear shapes can be replicated by both the normal and
tangential applied tension gradients, as well as by a variable bending rigidity.
As the force or modulus variation is increased, the imposed asymmetry produces
the transition from the non-polar bi-axial-symmetry (biconcave shape) to the
polar uniaxial symmetry (pear shape or kidney shape).  (2) The cell swimming
velocity is linear as a function of either the normal or tangential forces, but
the strong tension gradient in the tangential direction can further lead to a
kidney shapes and cause the swimming velocity to abruptly decrease. (3) The
velocity field within the cell forms two circulating loops which provide a
mechanism for shuttling actin monomer from the rear to the front, as has been
postulated by experimentalists, and is shown schematically in Figure
\ref{cortflow}.

Future work on 2D cells using the current model will focus on the interactions
between the different factors with a view toward more accurately replicating the
observed shapes and further investigating fluid effects such as differences in
the interior and exterior viscositites.  We believe that we will be able to obtain
most of the shapes that are observed with proper modification of the parameters
in the model. For example, some shapes probably arise via highly localized force
distributions, and can probably be obtained by suitable modification of the
applied tension distributions. Another aspect for study concerns movement in
confined spaces, which can be done with the present methods if the boundaries
are not too close to the cell, but which will require new methods when the cell
is in direct contact with the substrate.

Although the present study provides a first step for the modeling of
non-adhesive cell movement, several other questions deserve future consideration
in order to capture a more realistic picture of tension driven swimming. For
example,in the present work, we assumed a single layer integrating both effects
of membrane and cortical layers whereas real cells are endowed with one lipid
membrane layer and a separate  cortical layer, and a nucleus that makes  
deformations more difficult. It will be of great importance to include the two
separated layers and celluar nucleus in modeling for better comparison with
experimental systems. Another future issue is to study the environment effects
on cell migration, for example, what is the spatial confinement effect on cell
swimming velocities and trajectories and what is the cell dynamics in a 
viscoelastic fluid. The last issue is that in nature cells migrate in a 3D
environment, thus a more realistic model would incorporate the effect of the
second principal curvature on the migration phenomena of interest in this paper.
A 3D model would also allow for a more accurate description of the cortex-membrane
and fluid-membrane interactions. For example, applying an anisotropic force
distribution could lead to more realistic steady state shapes. A complete
investigation of this effect will be reported in future work. Some preliminary
results indicate that the cross section along the longest axis of the 3D shapes
adopted by cell membranes under tension coincide with the two dimensional
results shown here.  Figure  \ref{3Dshape} below shows the equilibrium shapes of a
membrane with and without an external tension gradient. 
 
\begin{figure}[h!]
\vspace*{-10pt}
 \hspace*{-10pt}  \includegraphics[width=2.0in]{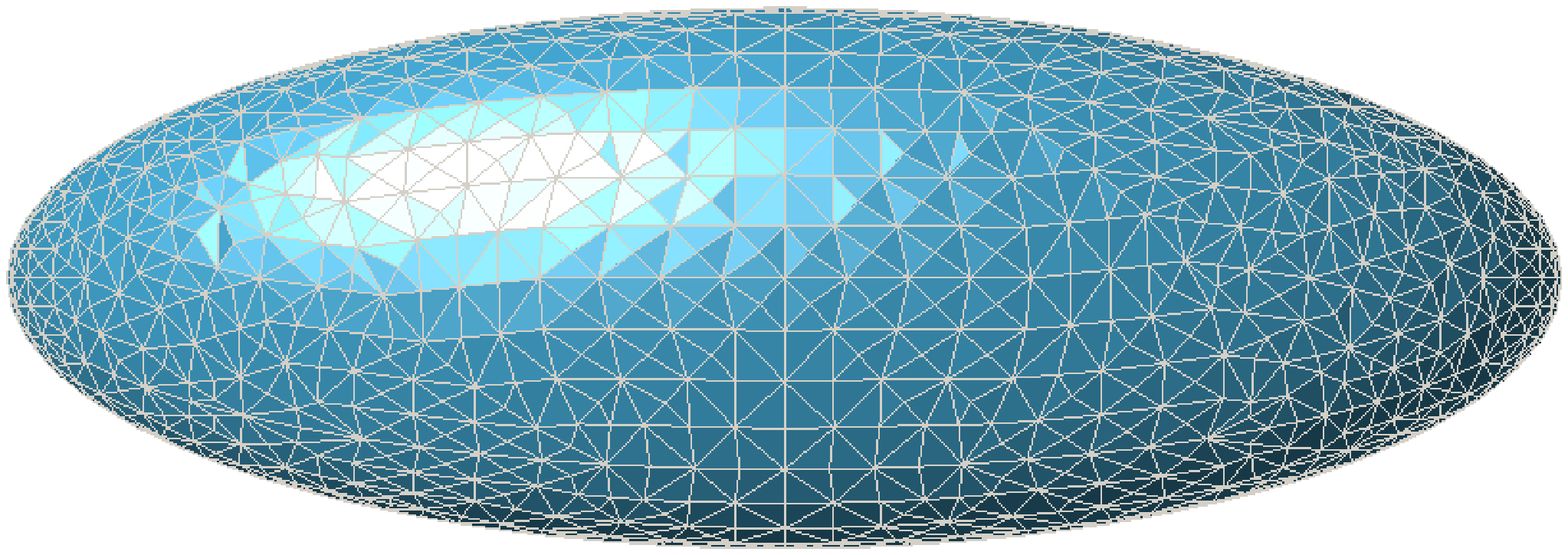} \hspace*{-5pt}\includegraphics[width=2.in]{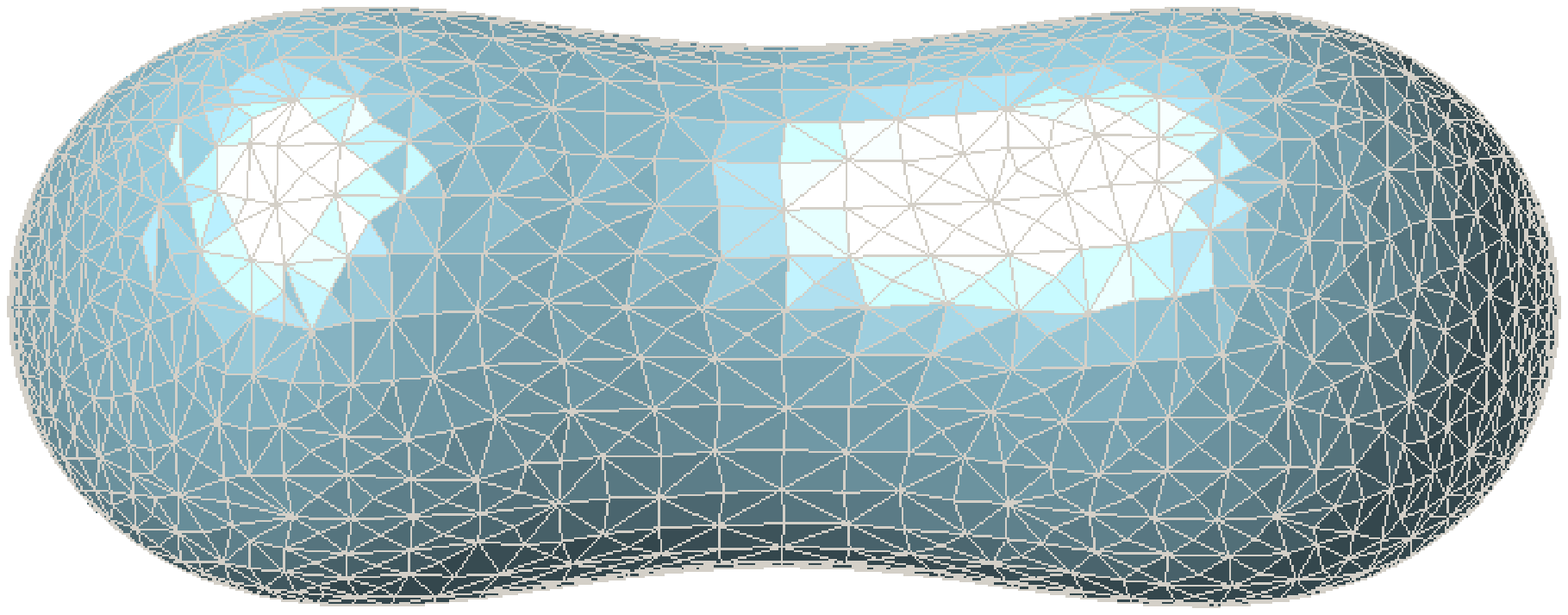}\hspace*{-5pt}\includegraphics[width=2.0in]{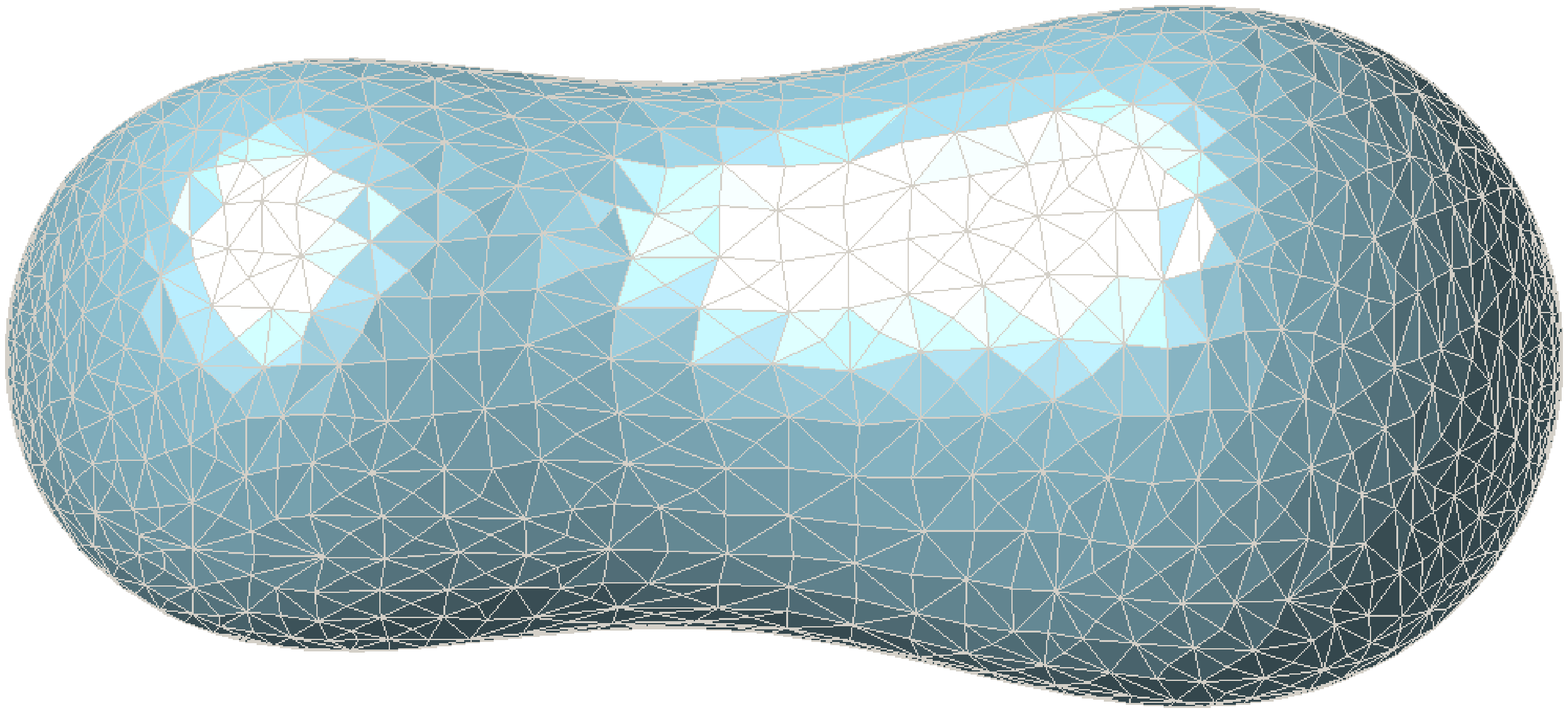}
 \caption{Here we exhibit the equilibrium shapes adopted by a cell membrane
  assumed to have an initial shape of a 3x1x1 ellipsoid shown at the 
  left. The figure in the center is the typical biconcave shape adopted by a
  membrane in the absence of cortical forces. At the right we can see the asymmetric
  final shape that is adopted when the membrane is under a linear tension gradient, going from  
  10pN/$\mu m^2$  in the back to 0 pN/$\mu m^2$ in the front. The numerical method
used to produce these shapes is based heavily on that of
\citep{Bonito:2010:PFG}. The computations were performed using the finite
element software FELICITY due to \cite{Walker:2017:FWD}.}  
\label{3Dshape}
\vspace*{-20pt}
\end{figure}
\section*{Acknowledgements}

We thank Prof. Misbah for generously providing a code written by Dr. Thi{\'e}baud and  used for the simulation of vesicle dynamics in an infinite channel \citep{Thiebaud:2013:RVS}.

%
\clearpage
\appendix 
\addcontentsline{toc}{section}{Appendices}
 \noindent{\bf \normalsize Appendices} \setcounter{equation}{0}
\renewcommand{\theequation}{\ref{Appendix A}.\arabic{equation}}

 \section{An outline of the derivation of the shape equation} 
\label{Appendix A}

In 1989 \citet{Ouyang:1989:BEV} computed the first, second and third order variations
of the Helfrich energy under normal deformations of  the membrane  by using the traditional
tensor analysis. Later  Capovilla {\it et al.}
\citep{Capovilla:2003:DGL} applied the covariant geometry and 
\citet{Tu:2004:GTE} applied Cartan's moving frame method to recalculate the
variation of the original Helfrich free energy in both normal and tangential
directions, respectively. Here, we extend these calculations to include
a heterogeneous bending and Gaussian modulus.
\begin{eqnarray}
\cF = \int_{\cS} \frac{1}{2}k_B\left(2H-C_0\right)^2\da +\int_{\cS}
k_G K\da +\int_{\cS}  \Lambda\da  +P\int_\Omega {\rm d}V 
\label{genericHelfrich}
\end{eqnarray}
 Note that the Gauss-Bonnet theorem is only valid for a constant Gaussian modulus, $k_G$.
 If $k_G$ is variable, then one cannot ignore the second term in the energy when performing the first order variation.  

For the derivation we introduce the following symbols and definitions.  The unit
normal vector $\mathbf{n}$ points outward.  $g$ is the surface metric
tensor, and $L$ is the extrinsic curvature tensor.  $\psi_{,j}
\equiv \partial_{j}\psi$, so
$\nabla_{i}\psi_{,j}=\psi_{,ij}-\Gamma^k_{ij}\psi_{,k}=\partial_{i}\psi_{,j}-\Gamma^k_{ij}\psi_{,k}$
where $\Gamma^k_{ij}$ is the Christoffel symbol of the second kind, which is
symmetric. $H = -(\kappa_1 + \kappa_2)/2$ is the mean curvature, $K =
\kappa_1\kappa_2$ is the Gaussian curvature, where $\kappa_1$ and $\kappa_2 $ are the
principal curvatures. The surface
Laplacian is ${\Delta}_s \equiv \nabla^2_s \equiv
(\sqrt{g})^{-1} \partial_i(\sqrt{g}g^{ij}\partial_j)$ and $\overline{\Delta}_s\,
=\overline{\nabla}^2_{s} \equiv
(\sqrt{g})^{-1} \partial_i(\sqrt{g}KL^{ij}\partial_j)$ and $i,j = 1,2$,
throughout.

We will use the following identities of first order variations taken in both the normal and tangential directions,
\begin{equation}
  \begin{cases}
     \delta\bfx= \phi^i\mathbf{e}_i+\psi\mathbf{n},  \\
     \delta\mathbf{e}_i= \left(\nabla_i\phi^k-\psi L_i^k\right)\mathbf{e}_k+\left(\phi^kL_{ki}+\psi_{,i}\right)\mathbf{n}, \\
     \delta\mathbf{n}= -\left(\phi^kL_k^i+\nabla^i\psi\right)\mathbf{e}_i, \\
     \delta g_{ij}=\nabla_i\phi_j+\nabla_j\phi_i-2\psi L_{ij}, \\
     \delta g^{ij}=-\nabla^i\phi^j-\nabla^j\phi^i+2\psi L^{ij}, \\
     \delta L_{ij}=\left(\nabla_i\nabla_j-2HL_{ij}+Kg_{ij}\right)\psi+L_{ik}\nabla_j\phi^k+L_{kj}\nabla_i\phi^k+\phi^k\nabla_kL_{ij},\\
     \delta g=g\left(2\nabla_i\phi^i-4H\psi\right), \\
     \delta \sqrt{g}=\sqrt{g}\left(\nabla_i\phi^i-2H\psi\right), \\
     \delta H=\left(2H^2-K\right)\psi+\frac{1}{2}\Delta\psi+\phi^k\nabla_kH, \\
     \delta K=\overline{\Delta}_s\psi +  2HK\psi+\phi^k\nabla_kK.
 \label{relations}
  \end{cases}
\end{equation}
%


 We can calculate the general variation of the modified Helfrich free energy as follows: 
\begin{eqnarray}
\delta \cF &=& \int_{\cS} \left[\left(\delta\sqrt{g}\right)\frac{1}{2}k_B\left(2H-C_0\right)^2+2\sqrt{g}k_B\left(2H-C_0\right)\left(\delta H\right)\right]\d2u\nonumber \\[5pt]
&+&\int_{\cS} \left[\left(\delta\sqrt{g}\right)k_GK+\sqrt{g}k_G\left(\delta K\right)\right]\d2u +\int_{\cS}  \left[\left(\delta\sqrt{g}\right)\Lambda\right]\d2u\nonumber \\[5pt]
&+&\frac{P}{3}\int_{\cS} \left[\left(\delta\sqrt{g}\right)\left({\bfx}\cdot{\mathbf{n}}\right)+\sqrt{g}\left(\delta {\bfx}\right)\cdot\mathbf{n}+\sqrt{g} \bfx\cdot\left(\delta\mathbf{n}\right)\right]\d2u \nonumber \\[5pt]
&=&\int_{\cS} \bigg\{\sqrt{g}\left(\nabla_i\phi^i-2H\psi\right)\frac{1}{2}k_B\left(2H-C_0\right)^2
+2\sqrt{g}k_B\left(2H-C_0\right)\left[\left(2H^2-K\right)\psi+\frac{1}{2}\Delta\psi+\phi^k\nabla_kH\right]\bigg\}\d2u\nonumber \\[5pt]
&+&\int_{\cS} \big[\sqrt{g}\left(\nabla_i\phi^i-2H\psi\right)k_GK+\sqrt{g}k_G\big(KL^{ij}\nabla_i\psi_{,j}+2HK\psi
+\phi^k\nabla_kK\big)\big]\d2u +\int_{\cS} \left[\sqrt{g}\left(\nabla_i\phi^i-2H\psi\right)\Lambda\right]\d2u\nonumber \\[5pt]
&+&\frac{P}{3}\int_{\cS}
\big\{\sqrt{g}\left(\nabla_i\phi^i-2H\psi\right)\left({\bfx}\cdot{\mathbf{n}}\right)+\sqrt{g}\left(\phi^i\mathbf{e}_i+\psi\mathbf{n}\right)\cdot\mathbf{n}-
\sqrt{g} \bfx\cdot\left[\left(\phi^kL_k^i+\nabla^i\psi\right)\mathbf{e}_i\right]\big\}\d2u \nonumber \\[5pt]
&=&\int_{\cS} \bigg\{\left(\nabla_i\phi^i-2H\psi\right)\frac{1}{2}k_B\left(2H-C_0\right)^2
+2k_B\left(2H-C_0\right)\left[\left(2H^2-K\right)\psi+\frac{1}{2}\Delta\psi+\phi^k\nabla_kH\right]\bigg\}\da\nonumber \\[5pt]
&+&\int_{\cS}  \big[\left(\nabla_i\phi^i-2H\psi\right)k_GK +k_G\big(KL^{ij}\nabla_i\psi_{,j}+2HK\psi+\phi^k\nabla_kK\big)\big] \da\nonumber \\[5pt]
&+&\int_{\cS}  \left[\left(\nabla_i\phi^i-2H\psi\right)\Lambda\right]\da+\frac{P}{3}\int_{\cS}  \big\{\left(\nabla_i\phi^i-2H\psi\right)\left({\bfx}\cdot{\mathbf{n}}\right)
+\left(\phi^i\mathbf{e}_i+\psi\mathbf{n}\right)\cdot\mathbf{n}- \bfx\cdot\left[\left(\phi^kL_k^i+\nabla^i\psi\right)\mathbf{e}_i\right]\big\}\da \nonumber \\[5pt]
&=&\int_{\cS}
\big\{\Delta\left[k_B\left(2H-C_0\right)\right]+k_B\left(2H-C_0\right)\left(2H^2+C_0H-2K\right)
+\overline{\Delta}\, k_G-2\Lambda H+P\big\}\psi\da\nonumber \\[5pt]
&-& \int_{\cS}  \bigg\{\frac{1}{2}\left(2H-C_0\right)^2\nabla_ik_B+K\nabla_ik_G
+\nabla_i\Lambda\bigg\}\phi^i\da
\label{finalfrst}
\end{eqnarray}
Thus, the two first functional derivatives (force components) in the normal and tangential directions are as follows:
\begin{eqnarray}
         F^n &=& -\frac{\delta \cF}{\delta\psi} = \big\{-\Delta_s\left[k_B\left(2H-C_0\right)\right]-k_B\left(2H-C_0\right)\left(2H^2+C_0H-2K\right) -\overline{\Delta}_s\, k_G+2\Lambda H-P\big\} \\[5pt]
\vspace*{15pt} 
     F^t_i &=& -\frac{\delta \cF}{\delta \phi^i} = \bigg\{\frac{1}{2}\left(2H-C_0\right)^2\nabla_i k_B+K\nabla_i k_G+\nabla_i\Lambda\bigg\}
 \label{tangen}
\end{eqnarray} 

 \section{Area and arc length conservation via an harmonic potential}  
\label{Appendix.B}

There are several ways in which the area conservation term in the
free energy (\ref{freeng2}) can be written, and here we show one which has
a simple interpretation in terms of springs between nodes and is equivalent to the
form given in (\ref{freeng2}) under an appropriate definition of $\Lambda$. This
method has been used by numerous authors
\citep{Henriksen:2004:MME,Finken:2008:TFV,Freund:2007:LMM,Wu:2015:AMC,Wu:2016:ASC},
and previous implementations of it have shown that the precision of
this method depends on the choice of the coefficient $\Lambda_0(u,v)$. The area
or arc-length can be conserved as precisely as desired by making
$\Lambda_0(u,v)$ large enough, with the disadvantage that this may 
require taking a small time step in the numerical algorithm.

The method is based on the following formulation of the area constraint,
\begin{eqnarray}
\frac{1}{2}\int  \frac{\Lambda_0(u,v)}{\sqrt{g_0}}
\left(\sqrt{g}-\sqrt{g_0}\right)^2 {\rm d}u{\rm d}v  = \frac{1}{2} \sum_{i=1}^{N} \left[\Lambda_i \frac{\left(A_i-A_0\right)^2}{A_0}\right]
\label{springlike}
\end{eqnarray}
where $\Lambda_0(u,v)$ is fixed in the
original reference frame. $g$ is the metric on the deformed surface and $g_0$
is the metric fixed in  the original surface. 

The equivalence between the spring-like penalty method and the Lagrange
multiplier method is as follows. The variation of the areal energy under a
perturbation can be written as 
\begin{eqnarray}
F_\Lambda\left({\mathbf{x}}+\epsilon{\mathbf{x}}_A\right)
&=&\int \frac{\Lambda_0(u,v)}{2|{\mathbf{x}}^0_u\times{\mathbf{x}}^0_v|} \big(|\left({\mathbf{x}}+\epsilon{\mathbf{x}}_A\right)_u\times\left({\mathbf{x}}+\epsilon{\mathbf{x}}_A\right)_v|
-|{\mathbf{x}}^0_u\times{\mathbf{x}}^0_v|\big)^2 {\rm d}u{\rm d}v
\label{purt}
\end{eqnarray}
Therefore the first variation can be written 
\begin{eqnarray}
\frac{\delta F_\Lambda}{\delta \epsilon}\bigg|_{\epsilon\rightarrow 0} &=&\int
\frac{\Lambda_0(u,v)}{|{\mathbf{x}}^0_u\times{\mathbf{x}}^0_v|}
\left(|\left({\mathbf{x}}+\epsilon{\mathbf{x}}_A\right)_u\times\left({\mathbf{x}}+\epsilon{\mathbf{x}}_A\right)_v|-|{\mathbf{x}}^0_u\times{\mathbf{x}}^0_v|\right)_{\epsilon\rightarrow  0}\nonumber \\[5pt] 
&\cdot&\frac{\left({\mathbf{x}}+\epsilon{\mathbf{x}}_A\right)_u\times\left({\mathbf{x}}+\epsilon{\mathbf{x}}_A\right)_v}{|\left({\mathbf{x}}+\epsilon{\mathbf{x}}_A\right)_u\times\left({\mathbf{x}}+\epsilon{\mathbf{x}}_A\right)_v|}\cdot\big[{\mathbf{x}}_{A,u}\times\left({\mathbf{x}}+\epsilon{\mathbf{x}}_A\right)_v 
+\left({\mathbf{x}}+\epsilon{\mathbf{x}}_A\right)_u\times{\mathbf{x}}_{A,v}\big]_{\epsilon\rightarrow
  0} {\rm d}u{\rm d}v\nonumber, \\ [5pt]  
&=&\int \Lambda_0(u,v)
\left(\frac{|{\mathbf{x}}_u\times{\mathbf{x}}_v|}{|{\mathbf{x}}^0_u\times{\mathbf{x}}^0_v|}-1\right)\frac{{\mathbf{x}}_u\times{\mathbf{x}}_v}{|{\mathbf{x}}_u\times{\mathbf{x}}_v|}\cdot\big({\mathbf{x}}_{A,u}\times{\mathbf{x}}_v
+ {\mathbf{x}}_u\times{\mathbf{x}}_{A,v}\big) {\rm d}u{\rm d}v\nonumber, \\ [5pt]
&=&\int \Lambda{\mathbf{n}}\cdot\left({\mathbf{x}}_{A,u}\times{\mathbf{x}}_v +{\mathbf{x}}_u\times{\mathbf{x}}_{A,v}\right) {\rm d}u{\rm d}v \nonumber, \\ [5pt]
&=&\int \Lambda\left[\left({\mathbf{x}}_v\times{\mathbf{n}}\right)\cdot{\mathbf{x}}_{A,u} +\left({\mathbf{n}}\times{\mathbf{x}}_u\right)\cdot{\mathbf{x}}_{A,v}\right] {\rm d}u{\rm d}v \nonumber, \\ [5pt]
&=&-\int \left\{\left[\Lambda\left({\mathbf{x}}_v\times{\mathbf{n}}\right)\right]_u+\left[\Lambda\left({\mathbf{n}}\times{\mathbf{x}}_u\right)\right]_v\right\}\cdot{\mathbf{x}}_{A} {\rm d}u{\rm d}v,
\label{firstder}
\end{eqnarray}
where 
$$\Lambda \equiv \Lambda_0(u,v)
\left(\frac{|{\mathbf{x}}_u\times{\mathbf{x}}_v|}{|{\mathbf{x}}^0_u\times{\mathbf{x}}^0_v|}-1\right)=\Lambda_0(u,v)
\left(\sqrt{\frac{g}{g_0}}-1\right).
$$
\noindent
By using the vector product formula
${\boldsymbol{a}}\times\left({\boldsymbol{b}}\times{\boldsymbol{c}}\right)=\left({\boldsymbol{a}}\cdot{\boldsymbol{c}}\right){\boldsymbol{b}}-\left({\boldsymbol{a}}\cdot{\boldsymbol{b}}\right){\boldsymbol{c}}$,
we obtain
\begin{eqnarray}
{\mathbf{x}}_v\times{\mathbf{n}}&=&{\mathbf{x}}_v\times\left(\frac{{\mathbf{x}}_u\times{\mathbf{x}}_v}{|{\mathbf{x}}_u\times{\mathbf{x}}_v|}\right) \nonumber, \\ [5pt]
&=&\frac{\left({\mathbf{x}}_v\cdot{\mathbf{x}}_v\right){\mathbf{x}}_u-\left({\mathbf{x}}_v\cdot{\mathbf{x}}_u\right){\mathbf{x}}_v}{|{\mathbf{x}}_u\times{\mathbf{x}}_v|} \nonumber, \\ [5pt]
&=&\frac{G{\mathbf{x}}_u-F{\mathbf{x}}_v}{\sqrt{EG-F^2}}.
\end{eqnarray}
Similarly, 
\begin{eqnarray}
{\mathbf{n}}\times{\mathbf{x}}_u&=&\left(\frac{{\mathbf{x}}_u\times{\mathbf{x}}_v}{|{\mathbf{x}}_u\times{\mathbf{x}}_v|}\right)\times{\mathbf{x}}_u \nonumber, \\ [5pt]
&=&\frac{\left({\mathbf{x}}_u\cdot{\mathbf{x}}_u\right){\mathbf{x}}_v-\left({\mathbf{x}}_u\cdot{\mathbf{x}}_v\right){\mathbf{x}}_u}{|{\mathbf{x}}_u\times{\mathbf{x}}_v|} \nonumber, \\ [5pt]
&=&\frac{E{\mathbf{x}}_v-F{\mathbf{x}}_u}{\sqrt{EG-F^2}},
\end{eqnarray}
where $E$, $F$ and $G$ are the coefficients of the  first fundamental form.

The surface divergence of a vector field is  
\begin{eqnarray}
\nabla_s\cdot{\mathbf{u}}^s= \frac{G{\mathbf{x}}_u-F{\mathbf{x}}_v}{|{\mathbf{x}}_u\times{\mathbf{x}}_v|}\cdot{\mathbf{u}}^s_u+\frac{E{\mathbf{x}}_v-F{\mathbf{x}}_u}{|{\mathbf{x}}_u\times{\mathbf{x}}_v|}\cdot{\mathbf{u}}^s_v
\end{eqnarray}
\citep{doCarmo:1976:DGC}, and the surface gradient of  a scalar field is 
\begin{eqnarray}
\nabla_s\Lambda=
\frac{G{\mathbf{x}}_u-F{\mathbf{x}}_v}{|{\mathbf{x}}_u\times{\mathbf{x}}_v|}\Lambda_u+\frac{E{\mathbf{x}}_v-F{\mathbf{x}}_u}{|{\mathbf{x}}_u\times{\mathbf{x}}_v|}\Lambda_v. 
\label{scal}
\end{eqnarray}
From the Weigarten equation,
${\mathbf{n}}_i=\nabla_i{\mathbf{n}}=-L^j_i{\mathbf{x}}_j$, we have that,
${\mathbf{x}}_v\times{\mathbf{n}}_u=L^u_u\left({\mathbf{x}}_u\times{\mathbf{x}}_v\right)$
and
${\mathbf{n}}_v\times{\mathbf{x}}_u=L^v_v\left({\mathbf{x}}_u\times{\mathbf{x}}_v\right)$, where we are assuming the Einstein summation convention. Because
the mean curvature $H=-{\rm Tr}\left(L^i_i\right)/2$, we can further obtain  
\begin{eqnarray}
{\mathbf{x}}_v\times{\mathbf{n}}_u+{\mathbf{n}}_v\times{\mathbf{x}}_u=-2H\left({\mathbf{x}}_u\times{\mathbf{x}}_v\right)=-2H{\mathbf{n}}|{\mathbf{x}}_u\times{\mathbf{x}}_v|\label{vect}.
\end{eqnarray}
From equation (\ref{scal}) and equation (\ref{vect}), we easily have
\begin{eqnarray}
\nabla_s\Lambda-2\Lambda H{\mathbf{n}}=\frac{\left[\Lambda\left({\mathbf{x}}_v\times{\mathbf{n}}\right)\right]_u+\left[\Lambda\left({\mathbf{n}}\times{\mathbf{x}}_u\right)\right]_v}{|{\mathbf{x}}_u\times{\mathbf{x}}_v|}.
\label{tenderiv}
\end{eqnarray}
Therefore, equation (\ref{firstder}) can be further calculated as
\begin{eqnarray}
\frac{\delta F_\Lambda}{\delta \epsilon}\bigg|_{\epsilon\rightarrow 0}
&=&-\int \left(\nabla_s\Lambda-2\Lambda H{\mathbf{n}}\right)\cdot{\mathbf{x}}_{A} |{\mathbf{x}}_u\times{\mathbf{x}}_v|{\rm d}u{\rm d}v \nonumber, \\ [5pt]
&=&-\int \left(\nabla_s\Lambda-2\Lambda H{\mathbf{n}}\right)\cdot{\mathbf{x}}_{A} \sqrt{g}{\rm d}u{\rm d}v \nonumber, \\ [5pt]
&=&-\int \left(\nabla_s\Lambda-2\Lambda H{\mathbf{n}}\right)\cdot{\mathbf{x}}_{A} {\rm d}A
\nonumber, \\ [5pt]
&=&-\int {\mathbf{f}}^{\rm T}\cdot{\mathbf{x}}_{A} {\rm d}A.
\label{firstder2}
\end{eqnarray}
Thus, the spring-like penalty method is equivalent to the Lagrange multiplier
method for the appropriate definition of $\Lambda$.

\section{A global area or volume conservation condition}
\label{Appendix C}

There is no Lagrange multiplier needed for the volume conservation in an
incompressible fluid -- the volume is automatically conserved if the
divergence-free condition is met. However, in practice there is still some
deflation in the process of numerical calculation. Thus a volume correction must
be executed at each step, in order to precisely conserve the initial volume.

The traditional formula, which consists of summing all the volumes of the
element triangular prisms has an obvious disadvantage in that it fails to
calculate the volume when the center of mass is sometimes located outside the
cell body. To avoid this disadvantage, we introduce the Minkowski identity for
the volume of an N-dimensional geometric object in the space $R^N$
\begin{eqnarray}
V_{\rm N}= \frac{1}{N}\int {\rm d} A_{\rm N} \left({\mathbf{x}}\cdot{\mathbf{n}}\right)= \frac{1}{N}\int {\rm d} r^{N-1} \left({\mathbf{x}}\cdot{\mathbf{n}}\right)
\label{Minkow}
\end{eqnarray}
where $N$ is the dimension of the geometric object. $V_{\rm N}$ and $A_{\rm N}$ are the volume and the area of the N-dimensional object.

To prevent the accumulation of errors of the area in 2D, or the volume in
3D, we implement a correction step due to
\citep{Freund:2007:LMM}. We state this correction for the volume case, the area version is
given in \citep{Freund:2007:LMM}.  Suppose we apply a small perturbation
$\mathbf{p}$ to the membrane surface. The volume can be
precisely conserved as follows. Consider the functional
\begin{eqnarray}
S_{\rm N}=\int \left(\mathbf{p}\cdot\mathbf{p}\right) {\rm d}r^{N-1}+\lambda\left[\frac{1}{N}\int \left(\mathbf{x}+\mathbf{p}\right)\cdot\mathbf{n} {\rm d}r^{N-1}-V_0\right],
\label{action}
\end{eqnarray}
where $N$ is the dimension of the geometric object and $\lambda$ is a Lagrange multiplier to keep the volume conserved. A shape which minimizes the action above will satisfy that
\begin{eqnarray}
\delta S_{\rm N}&=&\int \delta\mathbf{p} \left(2\mathbf{p}+\frac{\lambda \mathbf{n}}{N}\right) {\rm d}r^{N-1}+\delta\lambda\bigg[\frac{1}{N}\int \left(\mathbf{x}\cdot\mathbf{n}\right) {\rm d}r^{N-1} \nonumber \\
&+&\left(V_{\rm N}-V_0\right)\bigg]=0.
\label{minimizaction}
\end{eqnarray}
By solving equation (\ref{minimizaction}), we obtain
\begin{equation}
  \begin{cases}
     \mathbf{p}=-\frac{\lambda}{2N}\mathbf{n}   \\
     \\
     \lambda=\frac{2N^2\left(V_{\rm N}-V_0\right)}{A_{\rm N}}
 \label{solution}
  \end{cases}
\end{equation}
Thus, the corrected position vector to preserve the volume can be expressed as
\begin{eqnarray}
\mathbf{x}_{\rm crt}&=&\mathbf{x}+\mathbf{p} \nonumber \\
&=&\mathbf{x}-\frac{N\left(V_{\rm N}-V_0\right)}{A_{\rm N}}\mathbf{n}
\label{newposition}
\end{eqnarray}

Numerical checks show that that this method can preserve the volume to within $10^{-3}V_0$ .

\section{Discretization and parametrization  implementation of the  boundary integral
  method} 
\label{Appendix D}

The cell boundary curve is discretized into $N$ segments with $N$ points \citep{Wu:2015:AMC,Wu:2016:ASC}. We label the $N$ points in the boundary as $1,2,\cdots,k-1,k,k+1,\cdots,N$. The discrete form of surface force $\mathbf{f}_j^i$ (bending part and tension part) can be expressed using the procedures in Section \ref{QD2CS}. 

The velocity of each node on the cell membrane has the discrete form of the integral equation (\ref{bierewrite}) in terms of the trapezoid rule and can be approximated using a finite difference scheme
\begin{equation}
\mathbf{u}_j^k=\mathbf{u}_j^{\infty ,k}+\frac{1}{4 \pi}\sum_{i=1}^N [G]_{j}^{k i} \mathbf{f}_j^i[d s]_j^i,
\label{disStokeseq}
\end{equation} 
where $\mathbf{u}_j^{\infty ,k}$ is a constant velocity in the $k$-th node at
the $j$-th timestep and the bending part of $\mathbf{f}_j^i$ can be discretized
using equation (\ref{localcurv}). The tension part of $\mathbf{f}_j^i$ has been discretized
in Section \ref{QD2CS}. The standard discretization of the free space Green's
function $[G]_{j}^{k i}$ can be found in the book \citep{Pozrikidis:2002:PGB}. 

Assuming the step of evolution time is $\tau_0$, the time sequence as $\tau_0,2\tau_0,\cdots,(j-1)\tau_0,j\tau_0,(j+1)\tau_0,\cdots$. Equation (\ref{displace}) is transformed into 
\begin{equation}
\mathbf{r}_{j+1}^k-\mathbf{r}_{j}^k=\mathbf{u}_j^k \tau_0,
\label{dismotioneq}
\end{equation}

Finally, the new surface force on each node is estimated again from the new cell
configuration $\mathbf{r}_{j+1}^k$, and the whole scheme above is repeated for a
long time until a the stationary state is reached.

In this paper, the cell membrane contour is discretized using $N_{\rm m}=120$
nodes whose positions are updated at each time step of
$\tau_0=10^{-4}{\tau}_{\rm m}$. The relative errors corresponding to the area,
the perimeter and the reduced area are around $0.0008\%$, $0.035\%$, and
$0.0009\%$, respectively. The steady shape of the cell is assumed to be obtained
when the difference $|\mathbf{x}(\tau+10^5\tau_0)-\mathbf{x}(\tau)|$ between the
positions of the same node on the membrane at two different moments, separated
by $10^5$ timesteps is less than $10^{-5}$. The calculations are performed on a
cluster consisting of 24 Intel@Core i5 processors with $16$ GB RAM per
node. OpenMP directives are used to parallelize the matrix-vector product
computation. Each configuration of the steady shapes is completed via $10^7$
iterations. It is important to note that some of the cases reported in the phase
diagrams in results section ran over more than 6 hours on a 12-core node since
we decided to avoid using any cutoff or periodic boundary conditions in our
system, due to the long-range nature of the hydrodynamic interaction. The use of
the appropriate Green's function (imposing the velocity vanishes on the walls)
allowed us to avoid finite size effects, since we can literally consider an
infinite domain along the flow direction.

As a validation of our numerical algorithm, the full phase diagram of dynamics
of amoeboid swimming have been obtained by both the boundary integral method and the
immersed boundary method \citep{Wu:2015:AMC,Wu:2016:ASC}. The two methods give
the exact same results. In this paper, we explore the ranges of parameters that result
in interesting cell shapes, such as the the pear and kidney shapes.
In order to have a reference for the conversion of dimensionless units
into physical ones, the following dimensional numbers for cell blebs can be
used: $R_0=6\mu m$, $\mu = 10^{-3} {\rm Pa\cdot s}$ and $\kappa=10^{-18}
J$. This leads to a characteristic time of shape relaxation of about
${\tau}_{\rm m} \approx 0.2 s$, which is consistent with measured values
for some cells, such as the amoeboid cells in \citep{Arroyo:2012:REE}. 

\section{An accurate algorithm to determine the interior and exterior of a cell}
\label{Appendix E}

There are two existing algorithms, the ray casting algorithm and the winding
number algorithm, to judge whether one target point is located inside or outside
a 2D polygon. Here we propose a simpler but equally precise algorithm. The idea
is to use the fact that the sum of all the acute angles formed by the target
point and every pair of points that define a segment on the two dimensional
membrane contour is equal to $2\pi$ if the target point is inside the cell,
$\pi$ if the target point is on the smooth boundary and $0$ if the target point
is outside the cell. Here we assume that acute angles are calculated  by
uniting line elements successively in the clockwise direction. Because the
normal vector on the membrane points outward, if the cross prodcut of two
directional neighbor segments is negative, then the acute angle is positive,
vice versa. This 2D directional summation of the acute angles formed at the
target point is equivalent to calculate the winding number. But our method can
be easily generalized to the 3D case, the only change is that the criterion of
$2\pi$  rad has to be changed to $4\pi$  sr with a summation implemented on
all the solid angles formed by the target point and every three 
  nearest neighbor points on the triangularized 2D membrane surface.  In the 3D
case, ones also must pay attention to the directional segments of each
triangular element in order to determine the directional steradian. The winding
number algorithm cannot be extended to the 3D case. 


\addcontentsline{toc}{section}{References}
\bibliographystyle{jmb}
\bibliography{HaoWu,Shape,Motil,Comp7-23-17}

\begin{thebibliography}{}

\bibitem[Arroyo {\em et~al.}, 2012]{Arroyo:2012:REE}
Arroyo, M., Heltai, L., Mill{\'a}n, D., \& DeSimone, A. (2012).
\newblock {\em Proceedings of the National Academy of Sciences, } {\bf 109}
  (44), 17874--17879.

\bibitem[Barry \& Bretscher, 2010]{Barry:2010:DAN}
Barry, N.~P. \& Bretscher, M.~S. (2010).
\newblock {\em Proc. Natl. Acad. Sci.} {\bf 107} (25), 11376--11380.

\bibitem[Bergert {\em et~al.}, 2012]{Bergert:2012:CMC}
Bergert, M., Chandradoss, S.~D., Desai, R.~A., \& Paluch, E. (2012).
\newblock {\em Proc. Natl. Acad. Sci.} {\bf 109} (36), 14434--14439.

\bibitem[Bergert {\em et~al.}, 2015]{Bergert:2015:FTD}
Bergert, M., Erzberger, A., Desai, R.~A., Aspalter, I.~M., Oates, A.~C.,
  Charras, G., Salbreux, G., \& Paluch, E.~K. (2015).
\newblock {\em Nature cell biology, } {\bf }.

\bibitem[Binam{{\'e}} {\em et~al.}, 2010]{Biname:2010:WMC}
Binam{{\'e}}, F., Pawlak, G., Roux, P., \& Hibner, U. (2010).
\newblock {\em Molecular BioSystems, } {\bf 6} (4), 648--661.

\bibitem[Bonito {\em et~al.}, 2010]{Bonito:2010:PFG}
Bonito, A., Nochetto, R.~H., \& Pauletti, M.~S. (2010).
\newblock {\em Journal of Computational Physics, } {\bf 229} (9), 3171--3188.

\bibitem[Brailsford {\em et~al.}, 1980]{Brailsford1980}
Brailsford, J.~D., Korpman, R.~A., \& Bull, B.~S. (1980).
\newblock {\em Journal of theoretical biology, } {\bf 86} (3), 531--546.

\bibitem[Callan-Jones {\em et~al.}, 2016]{Callan-Jones:2016:CFS}
Callan-Jones, A., Ruprecht, V., Wieser, S., Heisenberg, C.-P., \& Voituriez, R.
  (2016).
\newblock {\em Physical review letters, } {\bf 116} (2), 028102.

\bibitem[Canham, 1970]{Canham:1970:MEB}
Canham, P.~B. (1970).
\newblock {\em J. Theor. Biol.} {\bf 26} (1), 61--76.

\bibitem[Capovilla {\em et~al.}, 2003]{Capovilla:2003:DGL}
Capovilla, R., Guven, J., \& Santiago, J. (2003).
\newblock {\em Journal of Physics A: Mathematical and General, } {\bf 36} (23),
  6281.

\bibitem[Charras \& Paluch, 2008]{Charras:2008:BLW}
Charras, G. \& Paluch, E. (2008).
\newblock {\em Nature Reviews Molecular Cell Biology, } {\bf 9} (9), 730--736.

\bibitem[Childress, 1981]{Childress:1981:MSF}
Childress, S. (1981).
\newblock {\em Mechanics of swimming and flying}, volume~2.
\newblock Cambridge: Cambridge Univ Pr.

\bibitem[Dai \& Sheetz, 1999]{Dai:1999:MTF}
Dai, J. \& Sheetz, M.~P. (1999).
\newblock {\em Biophys. J.} {\bf 77} (6), 3363--3370.

\bibitem[Dai {\em et~al.}, 1999]{Dai:1999:MCG}
Dai, J., Ting-Beall, H.~P., Hochmuth, R.~M., Sheetz, M.~P., \& Titus, M.~A.
  (1999).
\newblock {\em Biophys. J.} {\bf 77} (2), 1168--1176.

\bibitem[Diz-Mu{{\~n}}oz {\em et~al.}, 2016]{Diz-Munoz:2016:MTA}
Diz-Mu{{\~n}}oz, A., Thurley, K., Chintamen, S., Altschuler, S.~J., Wu, L.~F.,
  Fletcher, D.~A., \& Weiner, O.~D. (2016).
\newblock {\em PLoS Biol.} {\bf 14} (6), e1002474.

\bibitem[Diz-Mu{\~n}oz {\em et~al.}, 2010]{Diz-Munoz:2010:CDC}
Diz-Mu{\~n}oz, A., Krieg, M., Bergert, M., Ibarlucea-Benitez, I., Muller,
  D.~J., Paluch, E., \& Heisenberg, C.~P. (2010).
\newblock {\em PLoS Biology, } {\bf 8} (11), e1000544.

\bibitem[do~Carmo, 1976]{doCarmo:1976:DGC}
do~Carmo, M.~P. (1976).
\newblock {\em Differential geometry of curves and surfaces}.
\newblock Prentice Hall.

\bibitem[Evans \& Needham, 1987]{Evans:1987:PPS}
Evans, E. \& Needham, D. (1987).
\newblock {\em Journal of Physical Chemistry, } {\bf 91} (16), 4219--4228.

\bibitem[Farutin {\em et~al.}, 2014]{Farutin:2014:NSV}
Farutin, A., Biben, T., \& Misbah, C. (2014).
\newblock {\em Journal of Computational Physics, } {\bf 275}, 539--568.

\bibitem[Finken {\em et~al.}, 2008]{Finken:2008:TFV}
Finken, R., Lamura, A., Seifert, U., \& Gompper, G. (2008).
\newblock {\em The European Physical Journal E: Soft Matter and Biological
  Physics, } {\bf 25} (3), 309--321.

\bibitem[Freund, 2007]{Freund:2007:LMM}
Freund, J.~B. (2007).
\newblock {\em Physics of Fluids, } {\bf 19} (2), 023301.

\bibitem[Friedl \& Alexander, 2011]{Friedl:2011:CIM}
Friedl, P. \& Alexander, S. (2011).
\newblock {\em Cell, } {\bf 147} (5), 992--1009.

\bibitem[Friedl \& Wolf, 2010]{Friedl:2010:PCM}
Friedl, P. \& Wolf, K. (2010).
\newblock {\em The J. cell biology, } {\bf 188} (1), 11--19.

\bibitem[Ghigliotti {\em et~al.}, 2010]{Ghigliotti:2010:RDT}
Ghigliotti, G., Biben, T., \& Misbah, C. (2010).
\newblock {\em J. Fluid Mech.} {\bf 653}, 489--518.

\bibitem[Haeger {\em et~al.}, 2015]{Haeger:2015:CCM}
Haeger, A., Wolf, K., Zegers, M.~M., \& Friedl, P. (2015).
\newblock {\em Trends in cell biology, } {\bf 25} (9), 556--566.

\bibitem[Helfrich, 1973]{Helfrich:1973:EPL}
Helfrich, W. (1973).
\newblock {\em Zeitschrift f{\"u}r Naturforschung C, } {\bf 28} (11-12),
  693--703.

\bibitem[Henriksen \& Ipsen, 2004]{Henriksen:2004:MME}
Henriksen, J.~R. \& Ipsen, J.~H. (2004).
\newblock {\em The European Physical Journal E: Soft Matter and Biological
  Physics, } {\bf 14} (2), 149--167.

\bibitem[Hochmuth {\em et~al.}, 1996]{Hochmuth:1996:DFM}
Hochmuth, R., Shao, J., Dai, J., \& Scheetz, M. (1996).
\newblock {\em Biophys. J.} {\bf 70}, 358--369.

\bibitem[Howe {\em et~al.}, 2013]{Howe:2013:HDA}
Howe, J.~D., Barry, N.~P., \& Bretscher, M.~S. (2013).
\newblock {\em PloS One, } {\bf 8} (9), e74382.

\bibitem[Hung {\em et~al.}, 2016]{Hung:2016:CSS}
Hung, W.-C., Yang, J.~R., Yankaskas, C.~L., Wong, B.~S., Wu, P.-H.,
  Pardo-Pastor, C., Serra, S.~A., Chiang, M.-J., Gu, Z., Wirtz, D., {\em
  et~al.} (2016).
\newblock {\em Cell reports, } {\bf 15} (7), 1430--1441.

\bibitem[Joanny \& Prost, 2009]{Joanny:2009:AGD}
Joanny, J.-F. \& Prost, J. (2009).
\newblock {\em HFSP journal, } {\bf 3} (2), 94--104.

\bibitem[Kaoui {\em et~al.}, 2009]{Kaoui:2009:WRB}
Kaoui, B., Biros, G., \& Misbah, C. (2009).
\newblock {\em Phys. Rev. Lett.} {\bf 103} (18), 188101.

\bibitem[L{{\"a}}mmermann {\em et~al.}, 2008]{Lammermann:2008:RLM}
L{{\"a}}mmermann, T., Bader, B.~L., Monkley, S.~J., Worbs, T.,
  Wedlich-S{{\"o}}ldner, R., Hirsch, K., Keller, M., F{{\"o}}rster, R.,
  Critchley, D.~R., F{{\"a}}ssler, R., {\em et~al.} (2008).
\newblock {\em Nature, } {\bf 453} (7191), 51--55.

\bibitem[Lauga \& Davis, 2012]{Lauga:2012:VMP}
Lauga, E. \& Davis, A.~M. (2012).
\newblock {\em J. Fluid Mechanics, } {\bf 705}, 120--133.

\bibitem[Lee {\em et~al.}, 1990]{Lee:1990:DML}
Lee, J., Gustafsson, M., Magnusson, K.-E., \& Jacobson, K. (1990).
\newblock {\em Sci.} {\bf 247} (4947), 1229--1233.

\bibitem[Li {\em et~al.}, 2014]{Li:2014:PRB}
Li, X., Peng, Z., Lei, H., Dao, M., \& Karniadakis, G.~E. (2014).
\newblock {\em Phil. Trans. R. Soc. A, } {\bf 372} (2021), 20130389.

\bibitem[Lieber {\em et~al.}, 2015]{Lieber:2015:FRM}
Lieber, A.~D., Schweitzer, Y., Kozlov, M.~M., \& Keren, K. (2015).
\newblock {\em Biophys. J.} {\bf 108} (7), 1599--1603.

\bibitem[Lipowsky, 2014]{Lipowsky2014}
Lipowsky, R. (2014).
\newblock {\em Advances in colloid and interface science, } {\bf 208}, 14--24.

\bibitem[Liu {\em et~al.}, 2015]{Liu:2015:CLA}
Liu, Y.-J., Berre, M.~L., Lautenschlaeger, F., Maiuri, P., Callan-Jones, A.,
  Heuz{\'e}, M., Takaki, T., Voituriez, R., \& Piel, M. (2015).
\newblock {\em Cell, } {\bf 160} (4), 659--672.

\bibitem[Logue {\em et~al.}, 2015]{Logue:2015:ERA}
Logue, J.~S., Cartagena-Rivera, A.~X., Baird, M.~A., Davidson, M.~W., Chadwick,
  R.~S., \& Waterman, C.~M. (2015).
\newblock {\em Elife, } {\bf 4}, e08314.

\bibitem[Lorentzen {\em et~al.}, 2011]{Lorentzen:2011:ERR}
Lorentzen, A., Bamber, J., Sadok, A., Elson-Schwab, I., \& Marshall, C.~J.
  (2011).
\newblock {\em J. Cell Sci.} {\bf 124} (8), 1256--1267.

\bibitem[Luo {\em et~al.}, 2013]{Luo:2013:MMC}
Luo, T., Mohan, K., Iglesias, P.~A., \& Robinson, D.~N. (2013).
\newblock {\em Nature materials, } {\bf 12} (11), 1064--1071.

\bibitem[Maiuri {\em et~al.}, 2015]{Maiuri:2015:AFM}
Maiuri, P., Rupprecht, J.-F., Wieser, S., Ruprecht, V., B{{\'e}}nichou, O.,
  Carpi, N., Coppey, M., Beco, S.~D., Gov, N., Heisenberg, C.-P., {\em et~al.}
  (2015).
\newblock {\em Cell, } {\bf 161} (2), 374--386.

\bibitem[N{{\"u}}rnberg {\em et~al.}, 2011]{Nurnberg:2011:NAI}
N{{\"u}}rnberg, A., Kitzing, T., \& Grosse, R. (2011).
\newblock {\em Nature Reviews Cancer, } {\bf 11} (3), 177--187.

\bibitem[Ou-Yang \& Helfrich, 1989]{Ouyang:1989:BEV}
Ou-Yang, Z.-C. \& Helfrich, W. (1989).
\newblock {\em Physical Review A, } {\bf 39} (10), 5280.

\bibitem[Petrie \& Yamada, 2016]{Petrie2016multiple}
Petrie, R.~J. \& Yamada, K.~M. (2016).
\newblock {\em Current opinion in cell biology, } {\bf 42}, 7--12.

\bibitem[Pollard {\em et~al.}, 2000]{Pollard:2000:MMC}
Pollard, T.~D., Blanchoin, L., \& Mullins, R.~D. (2000).
\newblock {\em Annu Rev Biophys Biomol Struct, } {\bf 29} (1), 545--76.

\bibitem[Pozrikidis, 1992]{Pozrikidis:1992:BIS}
Pozrikidis, C. (1992).
\newblock {\em Boundary integral and singularity methods for linearized viscous
  flow}.
\newblock Cambridge University Press.

\bibitem[Pozrikidis, 2002]{Pozrikidis:2002:PGB}
Pozrikidis, C. (2002).
\newblock {\em A practical guide to boundary element methods with the software
  library BEMLIB}.
\newblock CRC Press.

\bibitem[Pozrikidis, 2003]{Pozrikidis:2003:MSC}
Pozrikidis, C. (2003).
\newblock {\em Modeling and simulation of capsules and biological cells}.
\newblock CRC Press.

\bibitem[Prost {\em et~al.}, 2015]{Prost:2015:AGP}
Prost, J., J{\"u}licher, F., \& Joanny, J. (2015).
\newblock {\em Nature Physics, } {\bf 11} (2), 111--117.

\bibitem[Rahimian {\em et~al.}, 2015]{Rahimian:2015:BIM}
Rahimian, A., Veerapaneni, S.~K., Zorin, D., \& Biros, G. (2015).
\newblock {\em J. Computational Physics, } {\bf 298}, 766--786.

\bibitem[Renkawitz {\em et~al.}, 2009]{Renkawitz:2009:AFT}
Renkawitz, J., Schumann, K., Weber, M., L{{\"a}}mmermann, T., Pflicke, H.,
  Piel, M., Polleux, J., Spatz, J.~P., \& Sixt, M. (2009).
\newblock {\em Nature cell biology, } {\bf 11} (12), 1438--1443.

\bibitem[Renkawitz \& Sixt, 2010]{Renkawitz:2010:MFG}
Renkawitz, J. \& Sixt, M. (2010).
\newblock {\em EMBO reports, } {\bf 11} (10), 744--750.

\bibitem[Ruprecht {\em et~al.}, 2015]{Ruprecht:2015:CCT}
Ruprecht, V., Wieser, S., Callan-Jones, A., Smutny, M., Morita, H., Sako, K.,
  Barone, V., Ritsch-Marte, M., Sixt, M., Voituriez, R., {\em et~al.} (2015).
\newblock {\em Cell, } {\bf 160} (4), 673--685.

\bibitem[Salbreux {\em et~al.}, 2012]{Salbreux:2012:ACM}
Salbreux, G., Charras, G., \& Paluch, E. (2012).
\newblock {\em Trends in Cell Biology, } {\bf 22} (10), 536--545.

\bibitem[Sanz-Moreno {\em et~al.}, 2008]{Sanz-Moreno:2008:RAI}
Sanz-Moreno, V., Gadea, G., Ahn, J., Paterson, H., Marra, P., Pinner, S.,
  Sahai, E., \& Marshall, C.~J. (2008).
\newblock {\em Cell, } {\bf 135} (3), 510--523.

\bibitem[Seifert, 1997]{Seifert:1997:CFM}
Seifert, U. (1997).
\newblock {\em Advances in Physics, } {\bf 46}, 13--137.

\bibitem[Seifert {\em et~al.}, 1991]{Seifert:1991:STV}
Seifert, U., Berndl, K., \& Lipowsky, R. (1991).
\newblock {\em Phys. Rev. A, } {\bf 44}, 1182--1202.

\bibitem[Thi{\'e}baud \& Misbah, 2013]{Thiebaud:2013:RVS}
Thi{\'e}baud, M. \& Misbah, C. (2013).
\newblock {\em Phys. Rev. E, } {\bf 88} (6), 062707.

\bibitem[Traynor \& Kay, 2007]{Traynor:2007:PRE}
Traynor, D. \& Kay, R.~R. (2007).
\newblock {\em Journal of Cell science, } {\bf 120} (Pt 14), 2318.

\bibitem[Tu \& Ou-Yang, 2004]{Tu:2004:GTE}
Tu, Z. \& Ou-Yang, Z.-C. (2004).
\newblock {\em Journal of Physics A: Mathematical and General, } {\bf 37} (47),
  11407.

\bibitem[van Zijl {\em et~al.}, 2011]{vanZijl:2011:ISM}
van Zijl, F., Krupitza, G., \& Mikulits, W. (2011).
\newblock {\em Mutation Research/Reviews in Mutation Research, } {\bf 728} (1),
  23--34.

\bibitem[Veerapaneni {\em et~al.}, 2009]{Veerapaneni:2009:NMF}
Veerapaneni, S.~K., Gueyffier, D., Biros, G., \& Zorin, D. (2009).
\newblock {\em J. Computational Physics, } {\bf 228} (19), 7233--7249.

\bibitem[Veerapaneni {\em et~al.}, 2011]{Veerapaneni:2011:FAS}
Veerapaneni, S.~K., Rahimian, A., Biros, G., \& Zorin, D. (2011).
\newblock {\em J. Computational Physics, } {\bf 230} (14), 5610--5634.

\bibitem[Walker, 2017]{Walker:2017:FWD}
Walker, S.~W. (2017).

\bibitem[Wang \& Othmer, 2015a]{Wang:2015:CAA}
Wang, Q. \& Othmer, H.~G. (2015a).
\newblock {\em J. Math. Biol.} {\bf 72}, 1893--1926.

\bibitem[Wang \& Othmer, 2015b]{Wang:2015:PDM}
Wang, Q. \& Othmer, H.~G. (2015b).
\newblock {\em Mathematical Biosciences and Engineering, } {\bf 12} (6),
  1303--1320.

\bibitem[Wang \& Othmer, 2016]{Wang:2016:AMM}
Wang, Q. \& Othmer, H.~G. (2016).
\newblock {\em arXiv preprint arXiv:1610.02090, } {\bf }.

\bibitem[Welch, 2015]{Welch:2015:CMF}
Welch, M.~D. (2015).
\newblock {\em Cell, } {\bf 160} (4), 581--582.

\bibitem[Winklbauer, 2015]{Winklbauer:2015:CAS}
Winklbauer, R. (2015).
\newblock {\em J. Cell Sci, } {\bf 128} (20), 3687--3693.

\bibitem[Wu {\em et~al.}, 2016]{Wu:2016:ASC}
Wu, H., Farutin, A., Hu, W.-F., Thi{\'e}baud, M., Rafa{\"\i}, S., Peyla, P.,
  Lai, M.-C., \& Misbah, C. (2016).
\newblock {\em Soft matter, } {\bf 12} (36), 7470--7484.

\bibitem[Wu \& Noguchi, 2013]{Wu2013effects}
Wu, H. \& Noguchi, H. (2013).
\newblock In: {\em AIP Conference Proceedings} volume 1518 pp. 649--653, AIP.

\bibitem[Wu {\em et~al.}, 2013]{Wu:2013:MPM}
Wu, H., Shiba, H., \& Noguchi, H. (2013).
\newblock {\em Soft Matter, } {\bf 9} (41), 9907--9917.

\bibitem[Wu {\em et~al.}, 2015]{Wu:2015:AMC}
Wu, H., Thi{\'e}baud, M., Hu, W.-F., Farutin, A., Rafa{\"\i}, S., Lai, M.-C.,
  Peyla, P., \& Misbah, C. (2015).
\newblock {\em Physical Review E, } {\bf 92} (5), 050701.

\bibitem[Wu \& Tu, 2009]{Wu:2009:TNI}
Wu, H. \& Tu, Z. (2009).
\newblock {\em The Journal of chemical physics, } {\bf 130} (4), 045103.

\bibitem[Yumura {\em et~al.}, 2012]{Yumura:2012:CDR}
Yumura, S., Itoh, G., Kikuta, Y., Kikuchi, T., Kitanishi-Yumura, T., \&
  Tsujioka, M. (2012).
\newblock {\em Biology open, } {\bf 2} (2), 200--209.

\bibitem[Zatulovskiy {\em et~al.}, 2014]{Zatulovskiy:2014:BDC}
Zatulovskiy, E., Tyson, R., Bretschneider, T., \& Kay, R.~R. (2014).
\newblock {\em The J. Cell Biology, } {\bf 204} (6), 1027--1044.

\bibitem[Zhao {\em et~al.}, 2010]{Zhao:2010:SBI}
Zhao, H., Isfahani, A.~H., Olson, L.~N., \& Freund, J.~B. (2010).
\newblock {\em J. Computational Physics, } {\bf 229} (10), 3726--3744.

\end{thebibliography}

\end{document}